\documentclass[times,twocolumn,final]{elsarticle}

\usepackage{medima}
\usepackage{framed,multirow}

\usepackage{amssymb}
\usepackage{amsmath}
\usepackage{bm}
\usepackage{latexsym}

\usepackage{url}
\usepackage{xcolor}

\usepackage{hyperref}
\usepackage{stfloats}

\usepackage[ruled,vlined]{algorithm2e}
\SetKwInput{kwInit}{Initialize}
\SetKwInput{kwRequ}{Require}
\SetKwInput{kwReturn}{Return}

\newtheorem{thm}{Theorem}[section]

\newtheorem{defn}{Definition}[section]

\newproof{pf}{Proof}

\definecolor{newcolor}{rgb}{.8,.349,.1}

\journal{Medical Image Analysis}

\begin{document}

\verso{Wenqi Huang \textit{et~al. }}

\begin{frontmatter}

\title{Deep Low-Rank Plus Sparse Network for Dynamic MR Imaging\tnoteref{codelink}}%
\tnotetext[codelink]{Code available: https://github.com/wenqihuang/LS-Net-Dynamic-MRI.}



\author[1,3]{Wenqi Huang\fnref{fn1}}

\author[1,3]{Ziwen Ke\fnref{fn1}}
\fntext[fn1]{Wenqi Huang and Ziwen Ke contributed equally to this manuscript}
\author[1]{Zhuo-Xu Cui}
\author[2]{Jing Cheng}
\author[2,3]{Zhilang Qiu}
\author[2]{Sen Jia}
\author[4]{Leslie Ying}
\author[2]{Yanjie Zhu}
\author[1,2,5]{Dong Liang\corref{cor1}}
\ead{dong.liang@siat.ac.cn}
\cortext[cor1]{Corresponding author.}

\address[1]{Research Center for Medical AI, Shenzhen Institutes of Advanced Technology, Chinese Academy of Sciences, Shenzhen, China}
\address[2]{Paul C. Lauterbur Research Center for Biomedical Imaging, Shenzhen Institutes of Advanced Technology, Chinese Academy of Sciences, Shenzhen, China}
\address[3]{Shenzhen College of Advanced Technology, University of Chinese Academy of Sciences, Shenzhen, China}
\address[4]{Department of Biomedical Engineering and Department of Electrical Engineering, The State University of New York, Buffalo, NY, USA}
\address[5]{Pazhou Lab, Guangzhou, China}

\received{xx Nov 2020}
\finalform{xx xxx 2021}
\accepted{16 July 2021}
\availableonline{xx xxx 2021}

\begin{abstract}
In dynamic magnetic resonance (MR) imaging, low-rank plus sparse (L+S) decomposition, or robust principal component analysis (PCA), has achieved stunning performance. However, the selection of the parameters of L+S is empirical, and the acceleration rate is limited, which are common failings of iterative compressed sensing MR imaging (CS-MRI) reconstruction methods. Many deep learning approaches have been proposed to address these issues, but few of them use a low-rank prior. In this paper, a model-based low-rank plus sparse network, dubbed L+S-Net, is proposed for dynamic MR reconstruction. In particular, we use an alternating linearized minimization method to solve the optimization problem with low-rank and sparse regularization. Learned soft singular value thresholding is introduced to ensure the clear separation of the L component and S component. Then, the iterative steps are unrolled into a network in which the regularization parameters are learnable. We prove that the proposed L+S-Net achieves global convergence under two standard assumptions. Experiments on retrospective and prospective cardiac cine datasets show that the proposed model outperforms state-of-the-art CS and existing deep learning methods and has great potential for extremely high acceleration factors (up to 24x).
\end{abstract}


\end{frontmatter}


\section{Introduction}
\label{sec1}
Dynamic magnetic resonance (MR) imaging plays a critical role in clinical application because of its ability to simultaneously reveal spatial structure and dynamic changes in the time dimension. The trade-off between spatial resolution and temporal resolution must be made due to the limited scan time in practice. To resolve this conflict, accelerating dynamic MR imaging (MRI) with a highly undersampled k-space has attracted great research interest.

Over the past decade, the application of compressed sensing (CS) has significantly increased MR imaging speed and efficiency because of its capability of reconstructing images from highly undersampled signals. The sparse prior is widely used in CS-based reconstruction methods. Studies in the early stage used fixed bases such as temporal Fourier transforms and wavelet transforms to sparsify images in the x-t domain \citep{2006_cs, 2007_sparsemri, otazo2010combination, ma2008efficient}. The proposal of dictionary learning and manifold learning pushed this research a step forward. In dictionary learning, the fixed basis was replaced by a learned adaptive basis from acquired data, which led to reconstruction performance superior to that of fixed basis methods \citep{2013_spatialtemporaldic, 2014_timesparsitydic, 2015_NLD}.
Most recently, advances in deep-learning-based methods have demonstrated great potential in fast MR imaging. Given enough data, deep learning methods will significantly improve the quality of the reconstructed image. These methods can be generally categorized as unrolling-based algorithms \citep{2017_DCCNN, 2018_CRNN, wang2019dimension,aggarwal2018modl} and those that are not unrolled \citep{wang2016accelerating, kwon2017parallel, han2018deep, wang2020deepcomplexmri}. These types have a much more relaxed constraint than traditional CS methods and dictionary learning/manifold learning methods. The methods that are not based on unrolling exploit a specific standard network to learn an end-to-end mapping from the undersampled k-space to the clean image, which usually requires a large amount of training data and extended training time. In unrolled methods, iterative CS reconstruction algorithms are unrolled into deep networks to learn the parameters and transforms in the reconstruction model, which means that such networks can perform well with a smaller training data size \citep{cheng2019model, liang2020deep}. Furthermore, the end-to-end mappings in the non-unrolled methods are less interpretable than the model-based ones. Therefore, unrolled networks are more prevalent in the MR reconstruction field \citep{2020_IRF-Net}.

Along with the development of the sparse prior, the low-rank prior was introduced as an extension of sparsity in two forms: low-rank and sparse (L\&S) \citep{lingala2011accelerated, zhao2012image} and low-rank plus sparse (L+S) \citep{candes2011robust, chandrasekaran2011rank, otazo2015low} methods. Previous works on L\&S aimed to find a solution that is both low rank and sparse, while the works on L+S decomposed the data matrix into a low-rank component ($L$) and a sparse component ($S$). Equivalent L+S decomposition, or robust principal component analysis (RPCA), is more natural for dynamic imaging because the L and S components can represent the slowly changing background and the dynamic foreground, respectively \citep{otazo2015low}. Works based on low-rank matrix completion have proven that dynamic MR images have a strong low-rank prior, which can be used to improve the reconstruction quality \citep{lustig2010calibrationless, gao2011robust, otazo2015low}. Because of its appropriate modeling, L+S has achieved successful application in many fields in addition to MR imaging, such as foreground and background separation in computer vision \citep{candes2011robust, mansour2014video}, clutter suppression in ultrasound flow imaging \citep{ashikuzzaman2019low}, and image alignment \citep{peng2012rasl}. However, the low-rank prior also brings more parameters to tune during reconstruction,
which complicates
the regularization parameters' empirical selection. In addition, the singular value decomposition performed in each iteration is time consuming, which runs against the speed requirement of fast imaging in practice.

In this paper, we aim to eliminate the drawbacks of the L+S method by unrolling it into a deep network. Here, we propose a deep low-rank-plus-sparse network (L+S-Net) for dynamic MRI reconstruction. First, we formulate the dynamic MR image as a low-rank plus sparse model under the CS framework. Then, an alternating linearized minimization method is adopted to solve the optimization problem. The recovery of the L component and the S component can be written in iterative form, in which the low-rank constraint corresponds to singular value thresholding (SVT). We unroll the tedious iterative steps into an $n$-block network and set $n$ as a relatively small integer, leading to a considerable reduction in reconstruction time. All the regularization parameters in the L+S model are set as learnable, including the learned singular value thresholding (LSVT), the update step size and the proximal operator, avoiding empirical selection.

The main contributions of this work can be summarized as follows: First, the idea of the learned low-rank + sparse prior is introduced to dynamic MR imaging for better reconstruction results. 
We designed an deep network with a learned SVT (LSVT) for exploiting the low-rank + sparse prior for dynamic MR imaging. Second, we provide a convergence analysis of our method, which is important but uncommon in recent MR reconstruction works. Compared to Momentum-Net \citep{2020_momentum-net}, the theoretical convergence analysis in our work has weaker assumptions, and the low-rank prior is included. Retrospective and prospective studies demonstrated that our method can reconstruct MR images with an extremely high acceleration rate on both single-coil and multi-coil data (up to 24x multi-coil), and the reconstruction time is acceptable  ($\leq$1.5s), which means the proposed approach has potential for low-latency cardiac MR imaging.


The remainder of the paper proceeds as follows: Section II provides the background and introduces the proposed methods. Section III summarizes the experimental details and the results to demonstrate the effectiveness of the proposed method, while the discussion and conclusions are presented in Section IV and Section V, respectively.

\section{Methodology}
\subsection{Problem formulation}
A typical linear imaging model for MRI can be written as
\begin{equation}
    AX = y,
\end{equation}
where $A: \mathbb{C}^N\rightarrow\mathbb{C}^M$ is the encoding matrix, $y\in \mathbb{C}^M$ is the undersampled k-space data measured during acquisition, and $X$ is the vectorized image to be reconstructed. The encoding matrix $A$ is determined by the acquisition protocol, which is already known. In single-coil imaging, the encoding matrix is $A = F_u$, where $F_u$ is a Fourier transform with undersampling. In regard to parallel imaging, $A = F_uS$, where $S$ denotes the coil sensitivities. In such scenarios, the problem of image reconstruction aims to recover a clean image $X$ from the undersampled data $y$. According to the Nyquist–Shannon sampling theorem, if the k-space data are so undersampled that they do not satisfy the Nyquist sampling criterion, reconstruction is difficult.
CS-based methods can successfully reconstruct images from $k$-space data that do not satisfy the sampling theorem by exploiting sparse priors (e.g., sparsity) of the signal. With the help of regularization terms, iterative optimization algorithms are often used to solve the inverse problem:
\begin{equation}
  \label{CSmodel}
  X^* = \arg \min_{X} \frac{1}{2}\| AX-y\|_2^2+\mathcal{R}(X), 
\end{equation}
where $\mathcal{R}(X)$ is a prior regularization term. Usually, the $l_1$-norm is used as the regularizer.

\subsection{L+S decomposition in dynamic MRI reconstruction}
\label{originalLSmodel}
In dynamic MRI, we usually formulate the image as a matrix instead of a vector. Each column of the image matrix represents a vectorized temporal frame. The L+S algorithm decomposes the image matrix $X$ as a superposition of the background component $L$ and the dynamic component $S$ \citep{otazo2015low}. Because each frame's background components have a strong correlation along the temporal dimension, the $L$ components, which are assumed to change slowly among frames, form a low-rank matrix. The dynamic component $S$, which is already sparser than the original image $X$ with the background suppressed, has a much sparser representation when a good sparse transform is performed. In optimization, the nuclear norm, which is the sum of the matrix's singular values, has been used as a good surrogate for minimizing the rank. Therefore, the optimization problem in Eq.~\eqref{CSmodel} can be written in L+S form:
\begin{equation}
  \label{LSmodel}
  \min_{L, S} \frac{1}{2}\| A(L+S)-y\|_2^2+\lambda_{L} \| L\|_* + \mathcal{R}(S),
\end{equation}
where $\mathcal{R}(S)$ represents the sparse regularizer and $\lambda_{L}$ is the regularization parameter of the low-rank term.

\subsection{The proposed method: L+S network for dynamic imaging}

To solve the optimization problem in Eq. \eqref{LSmodel}, we introduce an auxiliary variable $X$, which denotes the image to be reconstructed or the superposition of the $L$ component and the $S$ component, equivalently. The reformed optimization problem is as follows:
\begin{equation}
  \label{LSmodel_my}
  \min_{L, S, X} \frac{1}{2}\| AX-y\|_2^2+\lambda_{L} \| L\|_{*} + 
  \mathcal{R}(S), \quad s.t. \quad X = L+S.
\end{equation}
This problem can be solved via an alternating linearized minimization method to obtain the following subproblems (a detailed derivation can be found in Supplementary Material S1):
\begin{equation}
  \label{subproblemslong}
  \left\{
  \begin{aligned}
    L_{k+1} = \arg\min_{L} &\frac{\rho}{2}\|L+S_k-X_k\|_2^2 + \lambda_{L} \| L\|_{*}\\
    S_{k+1} = \arg\min_{S} &\frac{\rho}{2}\|L_{k+1}+S-X_k\|_2^2 + \mathcal{R}(S) \\
    X_{k+1} = \arg\min_{X} &\frac{\rho}{2}\|L_{k+1}+S_{k+1}-X\|_2^2 + 
              <\nabla F(L_{k+1} + S_{k+1}, X> \\
              &+ \frac{1}{2\eta}\|X-L_{k+1} - S_{k+1}\|_2^2\\
  \end{aligned}     
  \right.,
\end{equation}
where $F(X):= \frac{1}{2}\| Ax-y\|_2^2$ is the data fidelity term.
Each of the three subproblems can be considered a particular instance of the proximal gradient method. The subproblems can be solved by iterating between the following update steps:
\begin{equation}
  \label{updatesteps}
  \left\{
  \begin{aligned}
    L_{k+1} &= H_{\lambda_{L}}(X_k-S_k) \\
    S_{k+1} &= \mathcal{T}(X_k, L_{k+1}) \\
    X_{k+1} &= L_{k+1} + S_{k+1} - \gamma\nabla F(L_{k+1} + S_{k+1})
  \end{aligned}
  \right.,
\end{equation}
where
$\mathcal{T}(\cdot)$ is a proximal operator depending on the sparse regularizer $\mathcal{R}(\cdot)$; $\gamma$ is an update step size, where $\gamma = 1/(1+\eta\rho)$.
$H_{\lambda_L}(\cdot)$ is an SVT operator with a threshold of $\lambda_L$. Given that $M = U\Sigma V^*, \Sigma = \mathrm{diag}({\sigma_i}_{1\leq i \leq r})$ is the singular value decomposition of matrix $M$ (rank(M)=r), $H_{\lambda_L}(M)$ applies a soft threshold to the singular values of $M$ and yields
\begin{equation}
\label{SVT}
\begin{aligned}
    H_{\lambda_L}(M)&=U\mathcal{D}_{\lambda_L}(\Sigma)V^*,\\ \mathcal{D}_{\lambda_L}(\Sigma)&=\mathrm{diag}({\max(\sigma_i-\lambda_L, 0)}_{1\leq i \leq r}).
\end{aligned}
\end{equation}
$H_{\lambda_{L}}(\cdot)$ forces the matrix to be low-rank by eliminating small singular values with a threshold of $\lambda_{L}$. Because the last singular value is relatively small, some of the values are suppressed to zero, lowering the rank of $L$ as a consequence. The rank of $L$ is reduced at each iteration step, and finally, $L$ becomes a slowly changing background component. In traditional CS-MRI, the optimized $L^*$, $S^*$ and $X^*$ can be obtained by iteratively solving Eq.~\eqref{updatesteps}.

\begin{figure*}[t]
  \centering
  \includegraphics[width=\linewidth]{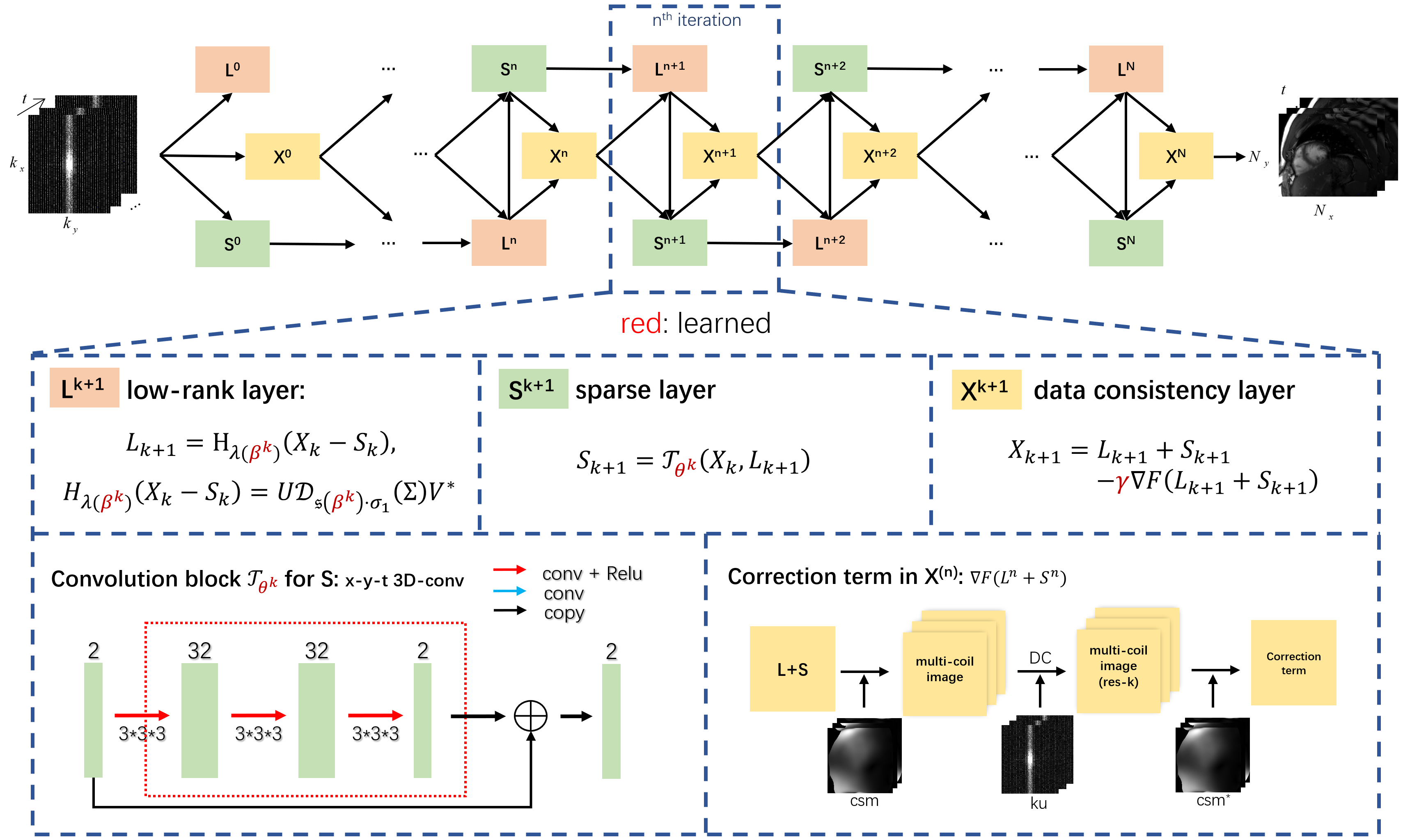}
\caption{\label{network}The proposed sparse plus low-rank network (L+S-Net) for dynamic MRI. L+S-Net is defined over the iterative procedures of Eq.~\eqref{networkmodules}. The three layers in Eq.~\eqref{networkmodules} correspond to the three modules in L+S-Net, which are named the low-rank prior layer $L^n$, sparse prior layer $S^n$ and data consistency layer $X^n$, respectively. The convolution block $C$ in the sparse prior layer $S^n$ is shown at the bottom left. The correction term $\nabla F(L^n+S^n)$ in the data consistency layer $X^n$ is shown at the bottom right.
}
\end{figure*}
Although the iterative steps are given, there are three intractable problems. First, because the proximal operator $\mathcal{T}(\cdot)$ is determined via the proximal gradient method, its closed form can only be obtained under the circumstance that the regularizer $\mathcal{R}$ is orthogonal (for example, as in a temporal fast Fourier transform). If a fixed orthogonal transform is adopted, the sparse representation will be limited. Otherwise, the calculation of $\mathcal{T}(\cdot)$ cannot be performed due to the lack of its closed form. Second, the selection of the parameters $\lambda_{L}, \gamma$ is empirical. There is no guidance for parameter selection; thus, the optimal parameters can only be found through repeated trials. This problem is severe when the reconstruction time is long. Third, the iteration scheme's algorithms usually take many steps to converge, making the reconstruction time very long.

The unrolling low-rank plus sparse network, dubbed L+S-Net, addresses all the problems well. In this study, the iteration steps are unrolled into several iteration blocks. Each of the blocks contains three network modules, as shown in Fig. \ref{network}, which are named the low-rank prior layer $L^k$, sparse prior layer $S^k$, and data consistency layer $X^k$, corresponding to the $L, S, X$ update steps in Eq.~\eqref{updatesteps}:
\begin{equation}
  \label{networkmodules}
  \left\{
  \begin{aligned}
    &L^{k+1}:&L_{k+1} &= H_{\lambda(\beta^k)}(X_k-S_k) \\ 
    &S^{k+1}:&S_{k+1} &= \mathcal{T}_{\theta^k}(X_k, L_{k+1}) \\
    &X^{k+1}:&X_{k+1} &= L_{k+1} + S_{k+1} - \gamma^k\nabla F(L_{k+1} + S_{k+1})
  \end{aligned}
  \right..
\end{equation}

{\bf A) The low-rank layer} $L^k$ obtains the learned soft threshold (LSVT) of $X_k-S_k$ according to the $L$ step in Eq.~\eqref{updatesteps}. This LSVT has the same calculation process as traditional SVT, as stated in Eq.~\eqref{SVT}, but the threshold $\lambda_{L}$ of the singular value thresholding operator $H_{\lambda_{L}}$ is replaced with a learnable threshold:
\begin{equation}
  H_{\lambda(\beta^k)}(M)=U\mathcal{D}_{\mathfrak{s}(\beta^k)\cdot\sigma_1}(\Sigma)V^*,
\end{equation}
where $\sigma_1$ is the maximal singular value in the singular value matrix $\Sigma$, and $\beta^k$ is the learnable threshold factor of the $k^{th}$ iteration block. The sigmoid function $\mathfrak{s}(\cdot)$ is employed to ensure that the threshold $\mathfrak{s}(\beta^k)\cdot\sigma_1$ is less than the maximal singular value and larger than zero. After performing LSVT, some small singular values are reduced to zero, which leads to a decrease in the rank.

{\bf B) The sparse layer} $S^k$ in Eq.~\eqref{networkmodules} uses a residual 3D convolutional neural network (CNN) $\mathcal{T}_{\theta^k}$ to learn a customized proximal operator for each unrolling block, taking the channel-stacked $X_k$ and $L_{k+1}$ as inputs. $\theta^k$ is the set of CNN parameters to be tuned in $S^k$. From the point of view of the mathematical interpretation of the original L+S methods, we call this module the sparse layer in this paper. However, this naming convention does not mean that the deep CNN that is learned is a sparse transform.
Like many other existing sparsity-driven deep learning methods, such as CRNN and DCCNN \citep{2018_CRNN,2017_DCCNN}, we learn a generalized sparse prior through the CNN. Therefore, the sparse regularizer is no longer limited to orthogonal regularizers, and the first problem is solved by learning a CNN for the proximal operator.

{\bf C) The data consistency layer} $X^k$:
In the update step for $X_{k}$, $\gamma^k$ is the learnable update step size for the $k^{th}$ iteration block. The gradient term is
\begin{equation}
  \nabla F(X) = A^*(A(X)-y),
\end{equation}
so the $X$ update step is a kind of data consistency operation, where $\nabla F(\hat{X}_{k+1})$ is the image correction term of the residual in k-space. $X^k$ has different forms in single-coil imaging and multicoil imaging. Specifically, for single-coil imaging,
\begin{equation}
  \label{singledc}
  \nabla F(X) = F_u^*(F_u(X)-y),
\end{equation}
where $F_u$ is a Fourier transform with undersampling and $F^*_u$ is the inverse Fourier transform. For multicoil imaging,
\begin{equation}
  \label{multidc}
  \nabla F(X) = S^*F_u^*(F_uS(X)-y),
\end{equation}
which adds the coil sensitivities $S$ into the calculation. The multicoil version is also illustrated in Fig. \ref{network}.


To address the parameter selection problem, the introduced parameters, including the threshold factor $\beta^k$, the update step $\gamma^k$ and the CNN proximal operator $\mathcal{T}_{\theta^k}$, are learned during the network training,
avoiding empirical manual selection.

For the problem of long reconstruction time, the unrolled network only contains several iteration blocks (usually less than 20), which corresponds to the several steps used in traditional iterative methods. The massive reduction in iteration steps (e.g., hundreds vs. several) drives the total reconstruction time down as a consequence.

These learnable parameters in the L+S-Net are different among the blocks, enabling the network to obtain the optimal reconstruction with only several blocks. The inputs $L_0$ and $X_0$ are initialized to a zero-filling image in the first unrolling block, and $S_0$ is initialized with zeros. The pseudocode of the L+S-Net algorithm is shown in Alg. \ref{alg}.

\begin{algorithm}[h]
\label{alg}
\SetAlgoLined
\kwRequ{$\{\mathcal{T}_{\theta^k}: k=0,...,N_{iter-1}\}, \{\gamma^k: k=0,...,N_{iter-1}\}, \{\beta^k: k=0,...,N_{iter-1}\}, y, A, F$}
\kwInit{$S_0=0, L_0=A^*y, X_0=A^*y$}
\For{$k=0,...,N_{iter-1}$}{
\textsc{Low-rank Prior:} \\
\qquad $L_{k+1} = H_{\beta^k}(X_k-S_k)$\;
\textsc{Sparse Prior:}\\
\qquad $S_{k+1} = \mathcal{T}_{\theta^k}(X_k, L_{k+1}) $\;
\textsc{Data Consistency:}\\
\qquad $X_{k+1} = L_{k+1} + S_{k+1} - \gamma\nabla F(L_{k+1} + S_{k+1})$\;
}
\kwReturn{$X_{N_{iter}}, L_{N_{iter}}, S_{N_{iter}}$}
\caption{L+S-Net}
\end{algorithm}

The formulated network should be trained before being put into use. The details of the training are given in Section \ref{modeltraining}.

Another pioneering work that integrates L+S into deep learning is CORONA \citep{2019_corona}, but it differs from our proposed approach in terms of application scenarios and technical solutions. 
CORONA was designed for clutter removal in ultrasound, while the proposed L+S-Net is used for dynamic MR imaging from undersampled k-space data. New application scenarios also bring new challenges, such as different data types (complex data), higher data dimensions (high-resolution \& multi-coil), different end goals of the application (L+S in L+S-Net vs L in CORONA), etc. In order to accurately reconstruct the phase information, we compute the real and imaginary parts as two channels in the convolutional network and keep the complex form in the rest of the computation in order to strike a balance between compatibility, computational speed and accuracy. We use coil sensitivity information to perform the conversion between multi-channel k-space data and channel-fused images, making high-quality multi-channel reconstruction possible. The proposed L+S-Net requires only fully-sampled images as labels, not separated L and S component images as CORONA does. In addition, the learned SVT in L+S-Net does not introduce new manually adjusted variables and is more convenient for training.

\subsection{Convergence analysis}
The convergence analysis of optimization algorithms is very important. In unrolling methods, the update steps are usually unfolded into a specific number of iteration blocks. It is also critical to study the convergence properties of these methods as $N_{iter}\rightarrow\infty$ \citep{2018_CT_convergence,2018_denoising_prior,2020_momentum-net}. If a given unrolling method does not tend to converge as $N_{iter}$ increases, the model has a high possibility of collapsing during training; in particular, the loss will fluctuate dramatically and may even
become a non-number value.
Nevertheless, only a few studies in deep learning-based MR reconstruction discuss the convergence property \citep{2020_momentum-net,2020_assembling_unrolling}. This section gives the basic assumptions and results for L+S-Net, and a detailed proof is attached in Supplementary Material S2.

\subsubsection{Preliminaries}
Before giving our results, we review some useful definitions.

We say $f$ is $L$-Lipschitz continuous; then, the inequality
$$|f(X)-f(Z)|\leq L\|X-Z\|$$
holds for any $X,Z$.
We say $f$ is $L$-smooth if the gradient of $f$ is $L$-Lipschitz continuous. An extended-real-valued function $f:dom(f)\rightarrow \mathbb{R}\cup\{+\infty\}$ is called proper if it is finite somewhere and never equals $-\infty$. We say $f$ is coercive if $f(X)\rightarrow\infty$ as $\|X\|\rightarrow\infty$.

\begin{defn}
[KL function]\label{kl} A proper function $f(\cdot)$ is called a Kurdyka-{\L}ojasiewicz (KL) function if any point $x$ where $f(\cdot)$ is subdifferentiable, i.e., $\partial f(X)\neq\varnothing$, satisfies the KL inequality.
That is, for all $X'$ that lie in the neighborhood of $X$ and satisfy $f(X)<f(X')<f(X)+\zeta$, the following inequality holds:
$$\varphi'(f(X')-f(X))\mathrm{dist}(0,\partial f(X'))\geq1,$$
where the scalar $\zeta\in[0,\infty)$; $\varphi:[0,\zeta)\rightarrow\mathbb{R}_{+}$ is a continuous concave function such that

\begin{itemize}
\setlength{\itemsep}{0pt}
\setlength{\parsep}{0pt}
\setlength{\parskip}{0pt}
\item $\varphi(0)=0$;
\item $\varphi$ is continuous differentiable in $(0,\zeta)$;
\item  for all $z\in(0,\zeta)$, $\varphi'(z)>0$.
\end{itemize}
\end{defn}

Since the network $\mathcal{T}(X, L)$ changes at each layer, the corresponding regularizers are also variable. Then, the penalty function for Eq. \ref{LSmodel_my} can be rewritten as follows:
\begin{equation}
\begin{aligned}
  \widetilde{J}_{\rho,k}(L,S):=&F(\widehat{X}_{k})+\langle \nabla F(\widehat{X}_{k}),L+S-\widehat{X}_{k}\rangle+\lambda_L\|L\|_*\\
  &+\mathcal{R}_{\theta^{k+1}}(S)+\frac{\rho}{2}\|L+S-\widehat{X}_{k}\|^2,
\end{aligned}
\end{equation}
where $\hat{X}_k=L_k+S_k$. Let $\mathcal{T}_{\theta^k}$ denote $\mathcal{T}$ at the $k^{th}$ layer; the alternating linearized minimization algorithm is depicted in Alg.  \ref{alg}.

\subsubsection{Assumptions}
Here are the two assumptions for the convergence analysis:

(\textbf{A}1) $\mathcal{R}_{\theta^1}:\mathbb{C}^d\rightarrow \mathbb{R}\cup\{+\infty\}$ is a nonnegative, proper, coercive and $L_{\mathcal{R}_{\theta^1}}$-smooth KL function.

(\textbf{A}2) The sequence of paired proximal operators of $(\mathcal{R}_{\theta^k},\mathcal{R}_{\theta^{k+1}} )$, termed $(\mathcal{T}_{\theta^k},\mathcal{T}_{\theta^{k+1}})$, is asymptotically nonexpansive with a sequence $\{\epsilon^{k+1}\}$; i.e.,
$$\|\mathcal{T}_{\theta^k}(X)-\mathcal{T}_{\theta^{k+1}}(Y)\|^2\leq(1+\epsilon^{k+1}) \|X-Y\|^2,~~\forall X,Y,k.$$

Assumptions (\textbf{A}1) and (\textbf{A}2) are standard assumptions for analyzing the convergence of learned iterative algorithms, which can also be found in \cite{2018_denoising_prior,2020_momentum-net}. Assumption (A2) is meant to ensure that the network $\mathcal{T}_{\theta^k}$ does not change too much at each layer.

Now, we define an auxiliary cost functional as follows:
\begin{equation*}
\begin{aligned}
    \widehat{J}_{\rho,k}(L,S):=&F(\widehat{X}_{k})+\langle \nabla F(\widehat{X}_{k}),L+S-\widehat{X}_{k}\rangle+\lambda_L\|L\|_*\\
    &+\mathcal{R}_{\theta^{1}}(S)+\frac{\rho}{2}\|L+S-\widehat{X}_{k}\|^2,
\end{aligned}
\end{equation*}
which is a linearized approximation of the following objective:
\begin{equation}
J(L,S):=F(L+S)+\lambda_L\|L\|_*+\mathcal{R}_{\theta^{1}}(S).
\end{equation}

\subsubsection{Convergence results}
This section gives the convergence results of L+S-Net. Given the assumptions mentioned before, we can prove that the reconstitution generated by L+S-Net satisfies the fixed-point condition.
\begin{thm}
\label{thm1}
Suppose that Assumptions (\textbf{A}1) and (\textbf{A}2) hold and that $\rho\geq\{L_{A^*A},\widetilde{\sigma}\}$, where $L_{A^*A}$ is the maximal singular value of $A^*A$ and $\widetilde{\sigma}$ is the minimum constant such that $\mathcal{R}_{\theta^{1}}(\cdot) +\frac{\widetilde{\sigma}}{2}\|\cdot\|^2$ is $\sigma$+4-strongly convex. Then, the sequence $\{(L_{k},S_{k})\}$ generated by Alg.  \ref{alg} converges to a critical point of the objective $J(L,S)$. The proof is given in the Supplementary Material.

\end{thm}

\section{Experimental results}
\subsection{Setup}
\subsubsection{Datasets}
Two cardiac datasets were used in this work, one for a retrospective study and another for a prospective study. The main experiments were conducted on the retrospective dataset to prove the effectiveness of the proposed method, and the prospective dataset was used to demonstrate the generalization performance. The proposed L+S-Net is capable of handling both single-coil and multicoil data. Therefore, the data were processed in both single-coil and multicoil schemes in this retrospective study.

1) Cardiac cine dataset for the retrospective study:

The fully sampled cardiac cine data were collected from 29 healthy volunteers on a 3T scanner (MAGNETOM Trio, Siemens Healthcare, Erlangen, Germany) with a multichannel receiver coil array (20 coils). All in vivo experiments were conducted with IRB approval and informed consent. For each subject, 10 to 13 short-axis slices were imaged with the retrospective electrocardiogram (ECG)-gated segmented bSSFP sequence during breath holding. A total of 386 slices were collected. The following sequence parameters were used: FOV $= 330 \times 330$ mm, acquisition matrix $= 256\times 256$, slice thickness = 6 mm, TR/TE = 3.0 ms/1.5 ms. The acquired temporal resolution was 40.0 ms, and each data point had approximately 25 phases that covered the entire cardiac cycle. We randomly selected 25 volunteers for training and the rest for testing. Deep learning typically requires a large amount of data for training; therefore, we applied data augmentation with stride and cropping. We slid a box of dimensions $192\times 192\times 18$ ($x\times y\times t$) in the dynamic images along the x, y, and t directions, and the stride along the three directions was 25, 25, and 7. Finally, we obtained 800 2D-t cardiac MR data of size 192×192×18 (x×y×t) for training and 118 data values for testing. Both single-coil and multicoil data were prepared for the experiments. The raw multicoil data of each frame were combined by the adaptive coil combining method \citep{walsh2000adaptive} to produce single-channel complex-valued images. The coil sensitivity maps were calculated for multicoil data from the undersampled time-averaged k-space center ($48\times 48$) using the ESPIRiT algorithm \citep{2014_ESPIRiT}.

2) Cardiac cine dataset for the prospective study:

\label{prospectivedata}
The prospective cardiac cine dataset contains two parts: 37 slices of full sampled training data for fine-tuning and 7 prospectively undersampled data for testing. Both the fully sampled and undersampled data were collected in short-axis and long-axis views on a 3T Siemens MAGNETOM Prisma machine using a bSSFP sequence with FOV=$800\times 300$ mm, acquisition matrix=$384\times 144$, slice thickness = 8 mm, and TR/TE = 38.4 ms/1.05 ms. The number of receiver coils is 34 for all data. The fully sampled data were collected with ECG gating and breath holding. The undersampled data were collected in real-time mode under free-breathing conditions. Because of the difference in the scan scheme, the ECG-gated fully sampled data have an average frame number of 23, and the prospective data have 65 frames for each case. All of these data are part of the OCMR dataset. Details can be found in \citep{chen2020ocmr}.

\subsubsection{Undersampling pattern}
Retrospective undersampling was performed to generate input/output pairs for network training or testing on the fully sampled data. For each frame, we fully sampled frequency encodes (along $k_x$) and undersampled phase encodes (along $k_y$). Both random Cartesian masks \citep{jung2007improved} and variable-density incoherent spatial-temporal acquisition (VISTA) masks \citep{ahmad2015variable} were used in this work to demonstrate generalization to different masks for our method. We applied a random Cartesian mask for the single-coil experiments, where it was ensured that 4 central phase encoders were sampled. The sampling mask for prospectively undersampled data is a 9-fold VISTA mask defined in the MR scanner.

\subsubsection{Model configuration}
\label{modeltraining}
The main structure of the network is shown in Fig. \ref{network}. In the experiments, L+S-Net is implemented with 10 iterative blocks. Each of them has independent learnable parameters and convolution layers. In the sparse prior blocks, each of the convolution blocks in $\mathcal{T}_{\theta^k}$ is a residual CNN with 3 convolution layers. Because the convolution is designed for floating-point data and is incapable of handling complex data, we divide each input of $\mathcal{T}_{\theta^k}$ into two channels, where the channels store the real and imaginary parts. The first two layers of the convolution block have 32 convolution kernels, and the last layer has 2 kernels to output the real and imaginary parts of the residual. The kernel size is $3\times3\times3$. Leaky rectified linear units (LeakyReLUs) \citep{2013_leakyReLU} are selected as the nonlinear activation functions after each convolution layer. $L_0$ and $X_0$ are initialized with a zero-filling image, and $S_0$ is initialized with zeros, as stated in Alg. \ref{alg}. The initial values for the learned threshold factor $\beta^k$ and the learned step size $\gamma^k$ in Eq.~\eqref{networkmodules} are -2 and 1, respectively.

The L+S-Net is trained in a supervised scheme. Given undersampled data as input and fully sampled images as the ground truth, the network is trained by minimizing the pixelwise mean-squared error (MSE) between the reconstructed image $X_{N_{iter}}$ and the labeled image $X_{Ref}$. The loss is as follows:
\begin{equation}
  \mathcal{L}(\{\theta^k,\beta^k,\gamma^k\}_{1\leq k\leq N_{iter}}) = \sum_M\sum_N\|X_{N_{iter}}-X_{Ref}\|_2^2
\end{equation}
where $\theta^k$ is the set of parameters in the CNN for $S^k$. $N$ is the pixel number of the reconstructed image, and $M$ is the number of training data.

A core part of the training is the backpropagation for parameters $\beta^k$ and $\gamma^k$.
Here, we present the key points of the gradient calculation:
\begin{equation}
    \begin{aligned}
    \frac{\partial L_{k+1}}{\partial \beta^k} &= U \tilde{\Sigma} V^*,\\
    \frac{\partial X_{k+1}}{\partial \gamma^k} &= -\nabla F(L_{k+1}+S_{k+1}),
    \end{aligned}
\end{equation}
where $X_k-S_k=U\Sigma V^*$ is singular value decomposition, and the diagonal matrix $\tilde{\Sigma}$ has
\begin{equation}
    \tilde{\Sigma}_{ii}=\left\{
    \begin{aligned}
    &-\mathfrak{s}'(\beta^k)\sigma_1, \quad  &\sigma_i-\mathfrak{s}(\beta^k)\sigma_1>0\\
    &0, &\sigma_i-\mathfrak{s}(\beta^k)\sigma_1\leq0
    \end{aligned}
    \right..
\end{equation}
Note that $\tilde{\Sigma}_{ii}$ denotes the (i, i)th entry of $\tilde{\Sigma}$, and $\sigma_1$ denotes the largest singular value in $\Sigma$. The gradient for the parameters in the whole network can be determined via the chain rule.

During the training, the batch size was 1 due to the limited GPU memory. An exponentially decaying learning rate \citep{2012_exponential_decay_lr} was used in the training procedure, and the initial learning rate was set to 0.001 with a decay of 0.95. The training procedure stopped after 50 epochs when the loss did not decrease any further.

The models were implemented in the open framework TensorFlow 2.2\citep{2016_tensorflow}. The training and testing were performed on a GPU server with an Nvidia RTX8000 graphics processing unit (GPU, 48 GB memory). The models were trained by the Adam optimizer \citep{kingma2014adam} with parameters $\beta_1 = 0.9$, $\beta_2 = 0.999$, and $\epsilon = 10^{-8}$. It took approximately 20 hours for 50 epochs to train the L+S-Net from scratch and approximately 2 hours for fine-tuning.

\begin{table}[b]
  \caption{\label{timetable}The supported data types, the number of model parameters, and the testing time for all comparison algorithms.}
  \centering
  \begin{tabular}{cccc}
  \hline\hline
   Method & Data & \#Param. & Time (s) \\
   \hline
   L+S & single/multi-coil & - & 40.12/94.56 \\
   DCCNN & single-coil & 432k & 0.43 \\
   CRNN & single-coil & 354k & 0.69 \\
   kt-NEXT & single-coil & 336k & 0.99 \\
   MoDL & multicoil & 339k & 0.80\\
   DL-ESPIRiT & multicoil & 334k & 0.38\\
   S-Net(ours) & single/multi-coil & 311k & 0.66/0.82 \\
   L+S-Net(ours) & single/multi-coil & 328k & 1.15/1.30\\

  \hline\hline
  \end{tabular}
\end{table}
\subsubsection{Performance evaluation}
\label{snet_def}
To determine the degree of improvement the learned low-rank prior brings to the model, we ablate the low-rank layers and formulate a sparse network (S-Net) for comparison. Since most existing works support either single-channel data or multichannel data, we set up two tracks for comparison. For the single-coil comparison, we chose L+S \citep{otazo2015low}, DCCNN \citep{2017_DCCNN}, CRNN \citep{2018_CRNN} and kt-NEXT \citep{2019_ktnext}. For the multicoil data, L+S, MoDL \citep{aggarwal2018modl} and DL-ESPIRiT \citep{2021_dl_espirit} were included. We did not compare with MoDL-SToRM (Biswas  et  al.,  2019) because that the SToRM prior needs the navigator signals to compute the manifold Laplacian matrix, while our data were acquired without a navigator. The differences between MoDL-SToRM and the proposed L+S-Net lie on two parts. Besides the navigator data, the SToRM prior $(\sqrt{w_{ij}\|x_i=x_j\|_p)^p}$ captures the similarity between different frames $x_i$ and $x_j$ in a low-dimension manifold, in which the weight $w_{ij}$ is learnable; while in the proposed L+S-Net, we combine a standard $\|L\|_*$ regularizer with deep learning to exploit the low-rankness in Casorati matrix of the dynamic image and the threshold factor in LSVT is learnable.

Table \ref{timetable} lists the supported data types, the number of model parameters, and the testing time for all comparison algorithms. We selected the data types as described in their original works. For a fair comparison, we adjusted the parameters of each CS-based method to achieve its best performance. All models of DL-based methods were tuned with a parameter size of approximately 330 K and the hyper-parameters were adjusted carefully to obtain
the best performance under the current dataset.

For the prospective reconstruction, the 8-fold acceleration models were fine-tuned for 20 epochs on the fully sampled data in the prospective dataset, which is described in Section 3.1.1(2). Real-time undersampled data were tested on these models in a prospective study.

Both visual comparison and quantitative evaluation were used for performance evaluation. For a quantitative evaluation, the mean squared error (MSE), peak signal-to-noise ratio (PSNR) and structural similarity index (SSIM) were calculated. A higher PSNR and SSIM and lower MSE indicate better quantitative performance.

For clarity, Table \ref{exp_list} lists all the experimental settings used in the following sections.
\begin{table}[t]
  \caption{\label{exp_list}The experimental arrangement for each section.}
  \centering
  \resizebox{\linewidth}{!}{
      \begin{tabular}{c|cccc}
      \hline\hline
      Section &\#Coils &Retro./Prosp. &AF &Mask	\\
      \hline
      3.2.1 &Single &Retrospective &8x &Random Cartesian \\
      3.2.2 &Single &Retrospective &8x, 10x, 12x &Random Cartesian \\
      3.2.3 &Multi &Retrospective &12x &VISTA \\
      3.2.4 &Multi &Prospective &9x &VISTA \\
      4.2 &Multi &Retrospective &16x, 20x, 24x &VISTA \\
      4.3 &Multi &Retrospective &12x &Random Cartesian \\
      \hline\hline
      \end{tabular}
  }
\end{table}

\subsection{The reconstruction performance of the proposed L+S-Net}
\subsubsection{The separation of the L and S components}
\label{lsseperation}
\begin{figure}[t]
  \centering
  \includegraphics[width=\linewidth]{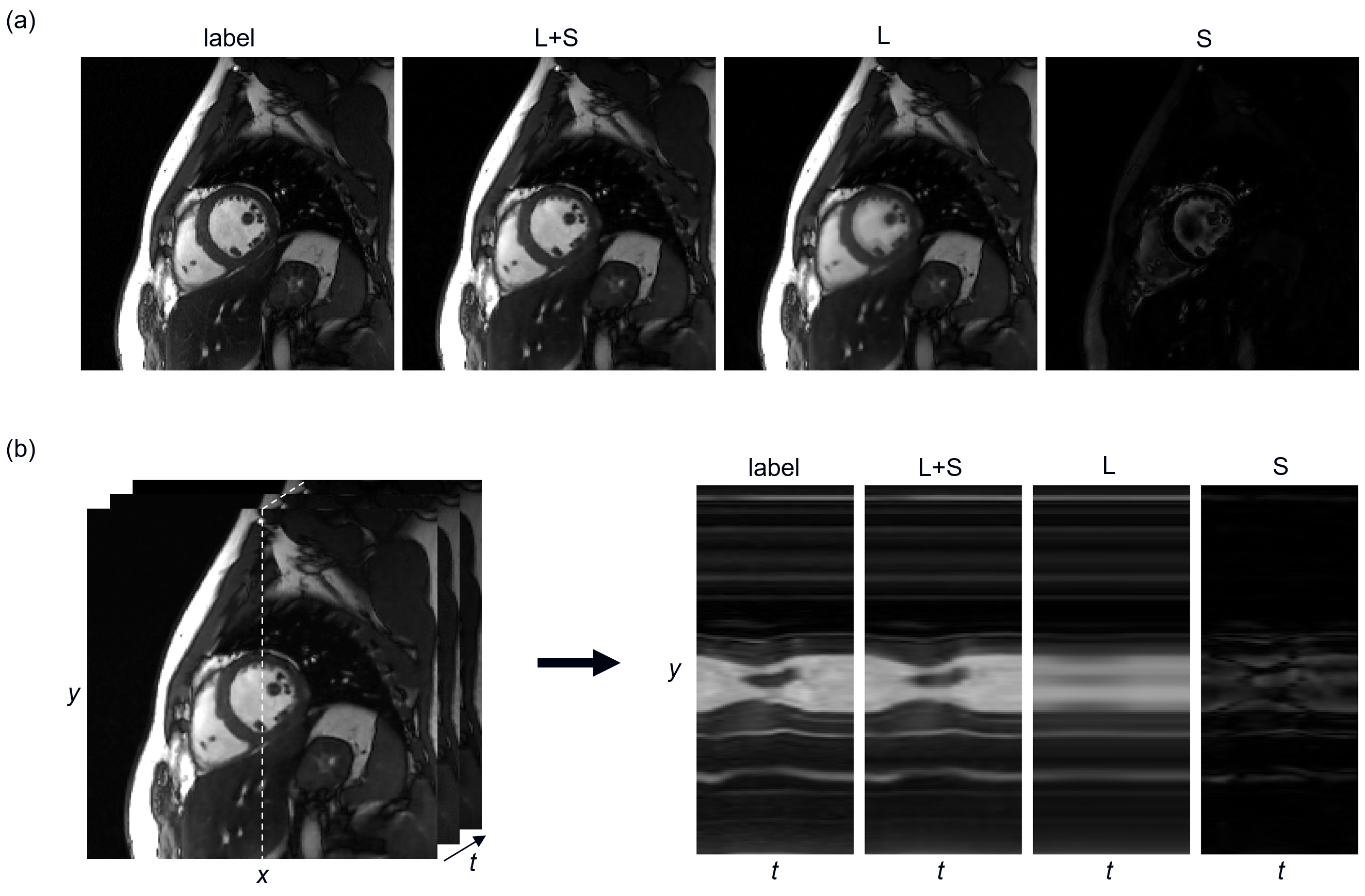}
\caption{\label{lscomponents}The $L$ and $S$ component separation results for the proposed L+S-Net with 8-fold acceleration. (a) The x-y view of the labeled image, the reconstruction result, and the reconstructed $L$ component and $S$ component. (b) The left part shows the results reconstructed by L+S-Net; the right part corresponds to the y-t view at the white centerline in the reconstructed image. The $L$ component changes slowly over time and contains the best-correlated cardiac motion component, rather than remaining static. The $S$ component is sparse even without a sparse transform.}
\end{figure}

The separation of the $L$ and $S$ components satisfies the hypothesis of being low-rank plus sparse. The reconstruction result $L+S$, together with its static component $L$ and dynamic component $S$, is shown in Fig. \ref{lscomponents}(a). The y-t view of the decomposition is shown in Fig. \ref{lscomponents}(b) to demonstrate the low-rank and sparse features. From the y-t view, we can clearly observe that the $L$ component is a slowly changing background rather than a static one. The low-rank property holds well in the test dataset. As can be seen from Fig.~\ref{lscomponents}, the S component is already sparse without a sparse transform.

\subsubsection{Comparison study for single-coil data}
\begin{figure}[!h]
  \centering
  \includegraphics[width=0.75\linewidth]{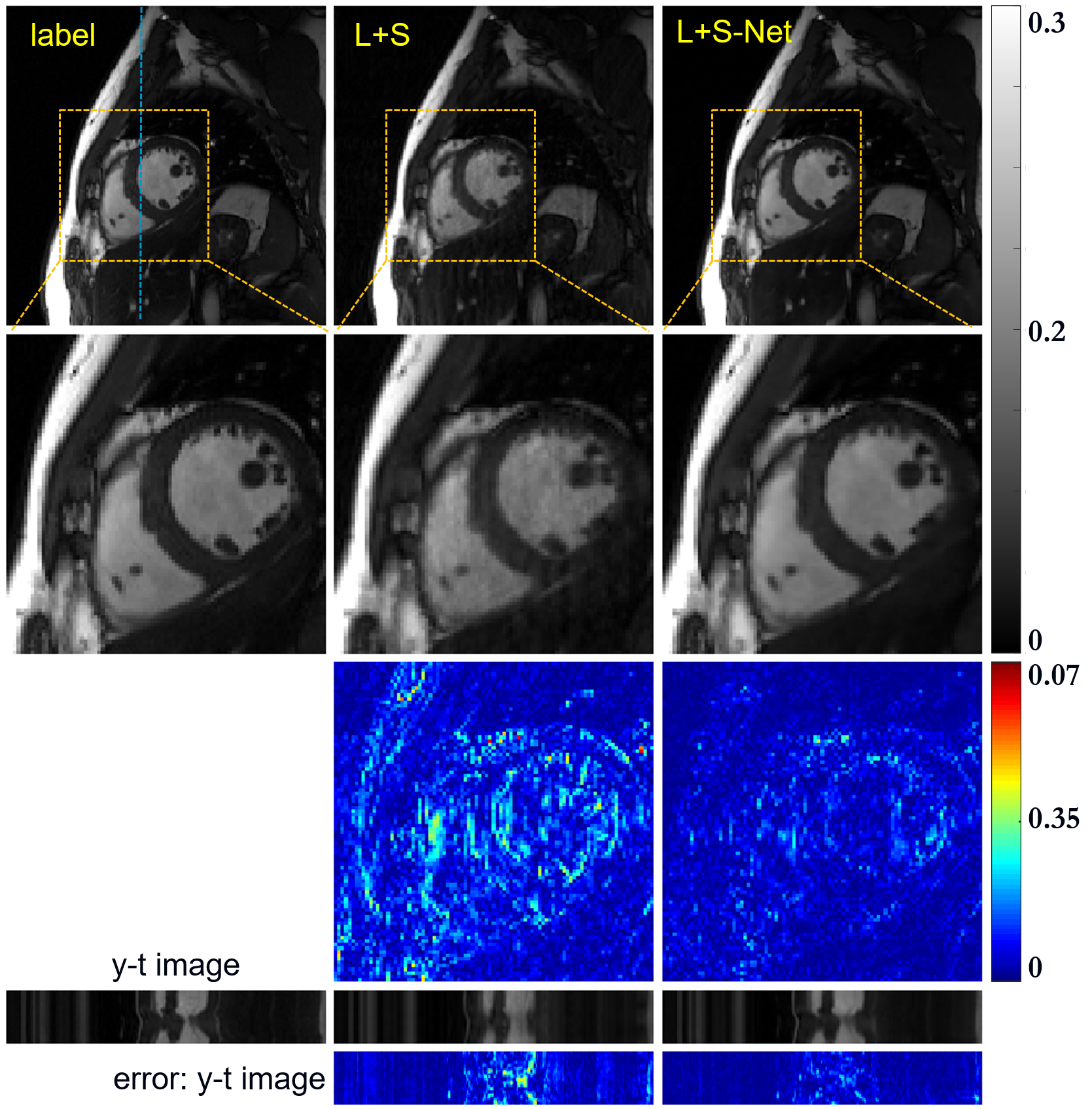}
\caption{\label{8xls-lsnet}The retrospective reconstruction results of L+S and L+S-Net at 8-fold acceleration on single-coil data. From left to right, the first row shows the ground truth and the reconstruction results of these methods. The second row shows the enlarged view of their respective heart regions framed by a yellow box. The third row shows the error map (display ranges [0, 0.07]). The y-t image (extraction of the 92nd slice along the y and temporal dimensions) and the y-t image error are also given for each signal to show the reconstruction performance in the temporal dimension.}
\end{figure}
\begin{figure}[!h]
  \centering
  \includegraphics[width=\linewidth]{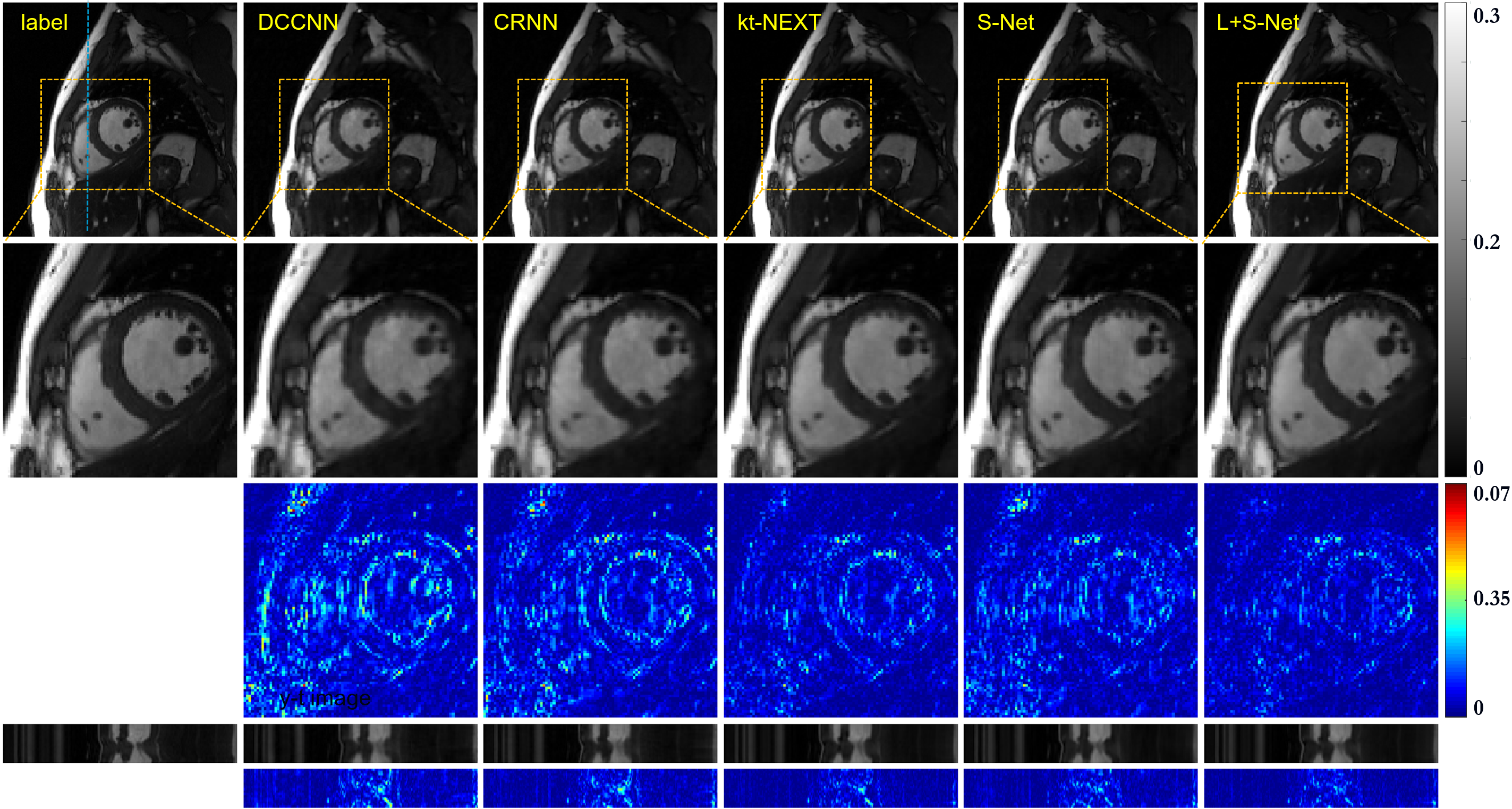}
\caption{\label{8xsota}The retrospective reconstruction results of the different methods (DC-CNN, CRNN, kt-NEXT, S-Net and L+S-Net) at 8-fold acceleration on single-coil data. From left to right, the first row shows the ground truth and the reconstruction results of these methods. The second row shows the enlarged view of their respective heart regions framed by a yellow box. The third row shows the error map (display range [0, 0.07]). The y-t image (extraction of the 92nd slice along the y and temporal dimensions) and the y-t image error are also given for each signal to show the reconstruction performance in the temporal dimension.
}
\end{figure}
\begin{figure*}[t]
  \centering
  \includegraphics[width=\linewidth]{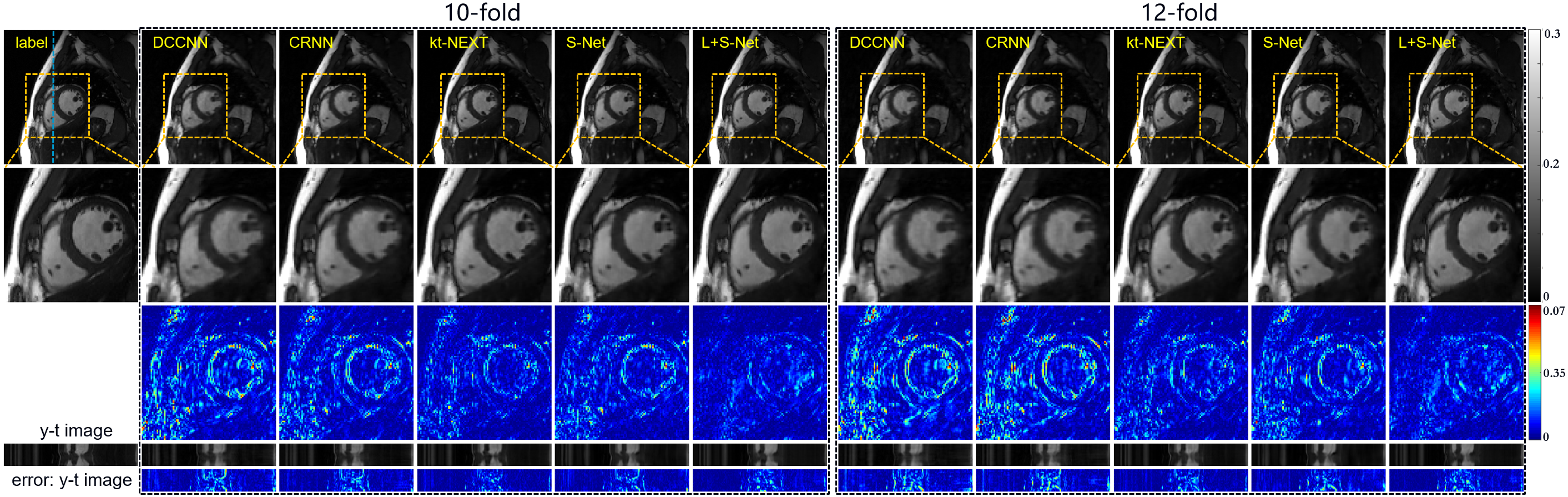}
\caption{\label{higheracc}Retrospective reconstruction results of the proposed L+S-Net and other methods at 10-fold and 12-fold acceleration on a single-coil cardiac cine dataset. The left half shows the 10-fold reconstruction, and the right half shows the 12-fold reconstruction. From left to right, the first row shows the ground truth and the reconstruction results of these methods. The second row shows the enlarged views of their respective heart regions framed by a yellow box. The third row shows the error map (display range [0, 0.07]). The y-t image (extraction of the 92nd slice along the y and temporal dimensions) and the y-t image error are also given for each signal to show the reconstruction performance in the temporal dimension.
}
\end{figure*}
\label{singlecompare}
The comparison between the unrolled L+S-Net and iterative L+S method at 8-fold acceleration is shown in Fig. \ref{8xls-lsnet}. The first row shows the ground truth and the reconstruction results for these two methods. The second row shows the magnified view of the heart region framed with a yellow box. The third row shows the error map of the ROI region. The y-t images at x=92 are shown in the fourth row, and the error maps of the y-t images are given in the last row. The display ranges for all greyscale images and error maps are $[0, 0.3]$ and $[0, 0.07]$, respectively. L+S-Net has an advantage over the traditional L+S method at 8-fold acceleration. The image reconstructed using L+S-Net appears much cleaner than that using the L+S method, while the L+S method exhibits artifacts within the chambers. The error of L+S is much larger than that of the proposed method, especially around the edge of the papillary muscle. From the y-t view, the proposed L+S-Net captures the dynamic information more precisely.
\begin{figure}[t]
  \centering
  \includegraphics[width=\linewidth]{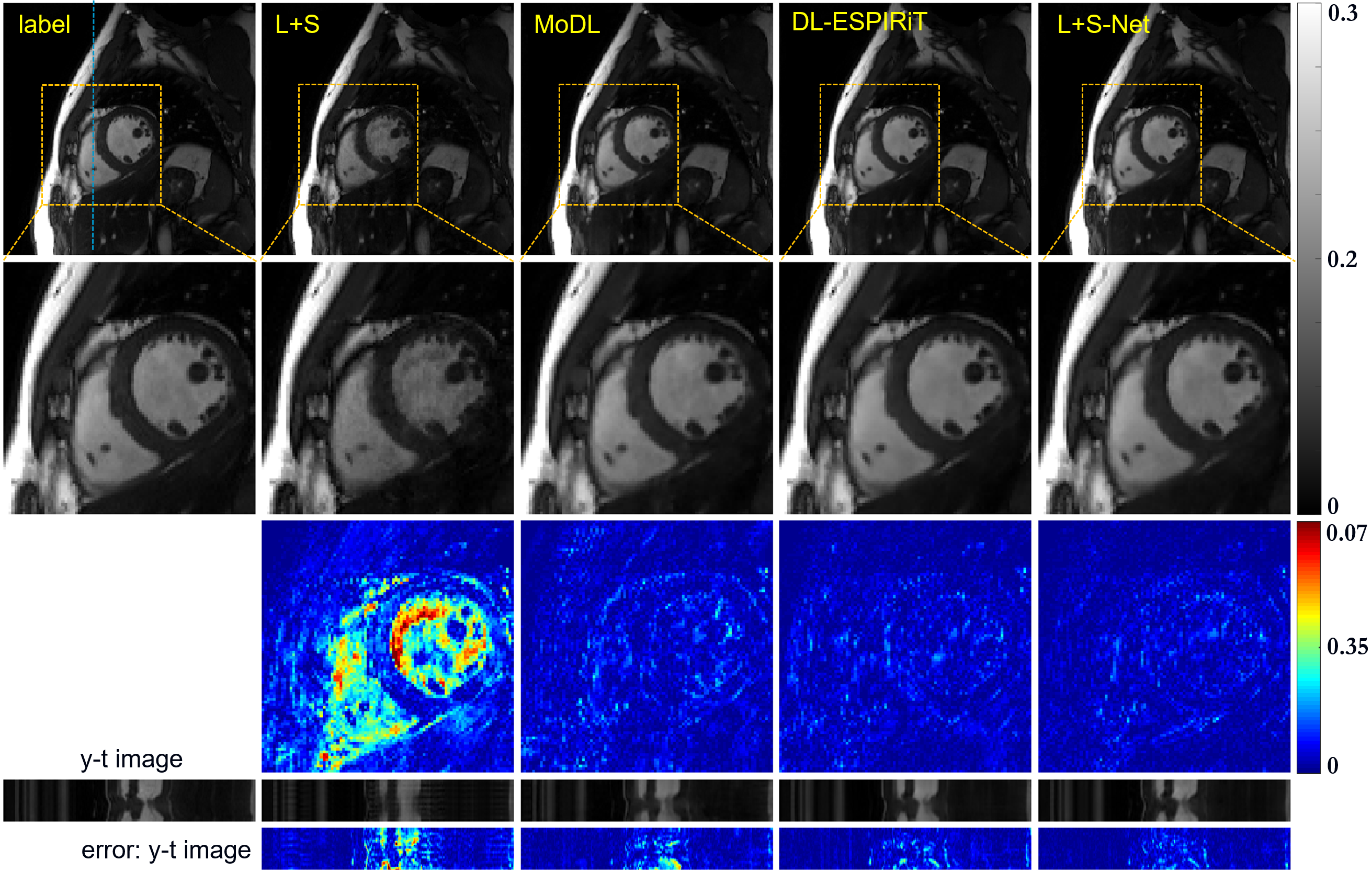}
\caption{\label{multicoil}
The retrospective reconstruction results of the different methods (L+S, MoDL, DL-ESPIRiT and L+S-Net) at 12-fold acceleration using multicoil data. From left to right, the first row shows the ground truth and the reconstruction results of these methods. The second row shows the enlarged views of their respective heart regions framed by a yellow box. The third row shows the error map (display range [0, 0.07]). The y-t image (extraction of the 92nd slice along the y and temporal dimensions) and the y-t image error are also given for each signal to show the reconstruction performance in the temporal dimension.}
\end{figure}
\begin{table}[t]
  \caption{\label{singlecoiltab}The average MSE, PSNR and SSIM of different methods at 8-fold, 10-fold and 12-fold acceleration on the single-coil cardiac cine test dataset (mean±std).}
  \centering
  \resizebox{\linewidth}{!}{
      \begin{tabular}{ccccc}
      \hline\hline
      AF&Methods &MSE(*e-5)	&PSNR	(dB)	&SSIM(*e-2)	\\
      \hline
      \multirow{6}{1em}{8x} &L+S &11.54±5.12	&39.75±1.75	&94.46±1.68\\
      &DCCNN	&7.43±2.33	&41.48±1.30	&96.22±0.76\\
      &CRNN	&5.60±1.67	&42.70±1.24	&97.07±0.61\\
      &kt-NEXT &3.33±1.38 & 45.10±1.65 &98.08±0.64\\
      &S-Net	&4.97±1.76	&43.28±1.47	&97.21±0.76\\
      &L+S-Net	&{\bf 2.91±1.25}	&{\bf 45.72±1.72}	&{\bf 98.23±0.61}\\
    
      \hline
      \multirow{6}{1em}{10x} &L+S &113.264±25.71	&29.57±0.97	&86.47±1.41\\
      &DCCNN	&11.07±3.13	&39.73±1.21	&94.61±0.87\\
      &CRNN	&7.28±2.25	&41.58±1.31	&96.49±0.75\\
      
      &kt-NEXT &5.04±1.73 & 43.21±1.40 &97.42±0.72\\
      &S-Net	&6.56±2.28	&42.08±1.47	&96.49±0.91\\
      &L+S-Net	&{\bf 4.39±1.60}	&{\bf 43.83±1.50}	&{\bf 97.68±0.68}\\ 
      \hline 
      \multirow{6}{1em}{12x}&L+S &113.42±25.90	&29.56±0.99	&86.46±1.45\\
      &DCCNN	&12.98±3.62	&39.03±1.19	&93.78±0.87\\
      &CRNN	&11.87±3.35	&39.43±1.21	&94.57±0.89\\
      
      &kt-NEXT &5.72±2.32 & 42.74±1.60 &96.99±0.92\\
      &S-Net	&7.70±2.55	&41.30±1.40	&95.99±0.97\\
      &L+S-Net	&{\bf 4.94±1.89}	&{\bf 43.35±1.57}	&{\bf 97.40±0.77}\\
    
      \hline\hline
      \end{tabular}
  }
\end{table}

Our proposed method also exhibits better performance than the state-of-the-art reconstruction methods. Fig.~\ref{8xsota} depicts the reconstruction results at 8-fold acceleration for the cardiac cine dataset. The proposed method outperforms DCCNN, CRNN and kt-NEXT, which can be clearly seen from the error maps. The comparison between these methods shows that L+S-Net performs better than the other methods in detail retention and artefact removal. The comparison between L+S-Net and S-Net (the same network structure as L+S-Net without the low-rank prior layers, described in Section \ref{snet_def}) shows that, with the low-rank prior, L+S-Net exhibits better performance in detail reconstruction and contrast, which leads to a lower error level around edges and high-frequency areas. This result indicates that the low-rank prior is critical for improving dynamic MR reconstruction.

The quantitative evaluation is shown in Table \ref{singlecoiltab}. Corresponding to the visual comparison result, L+S-Net achieves the best quantitative performance. Therefore, the proposed L+S-Net can effectively explore low-rank priors through learning and consequently improve the reconstruction quality.

The proposed L+S-Net uses both a sparse prior and a low-rank prior in dynamic images. Using more regularization terms in the optimization problem greatly improves the reconstruction performance, making a higher acceleration factor possible. We tested our proposed L+S-Net with a higher undersampling rate at 10-fold and 12-fold acceleration on the cardiac cine dataset to show its potential in highly undersampled imaging. Fig.~\ref{higheracc} shows the 10-fold and 12-fold accelerated reconstruction results. The proposed L+S-Net shows superior performance. At 10-fold acceleration, L+S-Net outputs images with good anatomical details of the heart, and the papillary muscle near the boundary is still clear. At 12-fold acceleration, the reconstructed image is still acceptable. Most of the details, such as the papillary muscle near the boundary, are still noticeable, although high-frequency details in the images, mainly around the edge of the cardiac muscle, are slightly blurry. The quantitative metrics are given in Table \ref{singlecoiltab}, which shows that the proposed L+S-Net still has good performance at a higher acceleration factor.

\begin{figure*}[t]
  \centering
  \includegraphics[width=\linewidth]{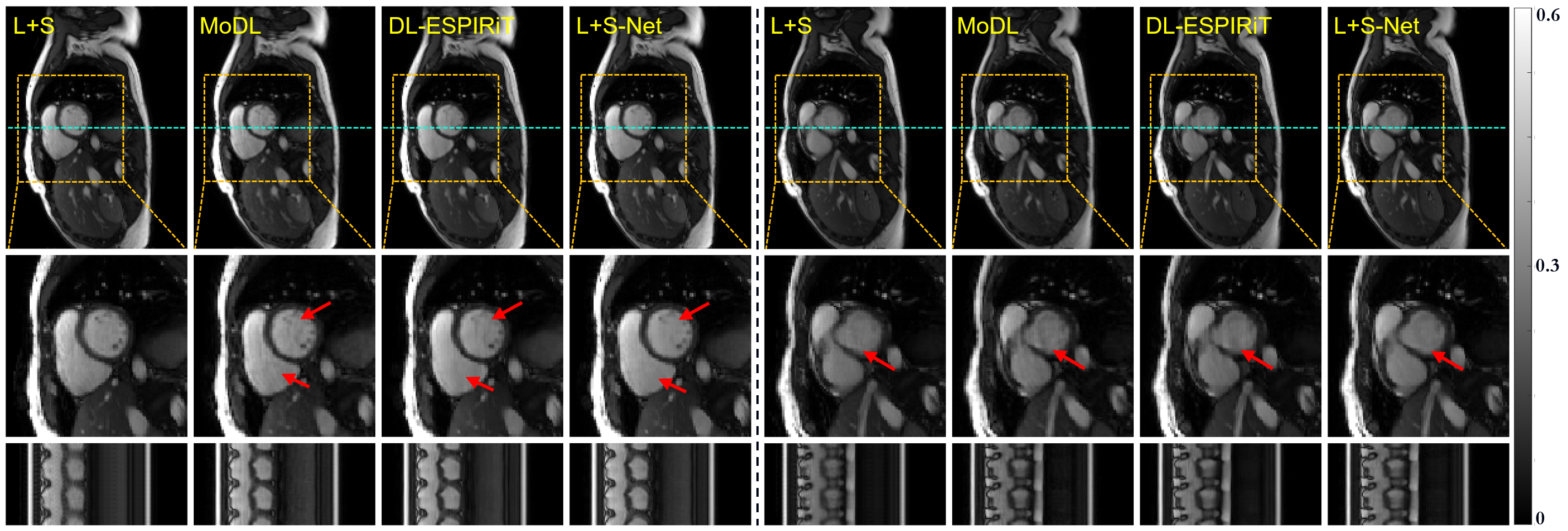}
\caption{\label{prospectivefig}The prospective reconstruction results of L+S and the proposed L+S-Net at 9-fold acceleration. The left half and the right half show 2 different test cases. The first row shows the whole image at t=18. The second row shows the enlarged views of their respective heart regions framed by a yellow box. The third row shows the y-t images extracted from the white lines in the first row of images.}
\end{figure*}

\subsubsection{Multicoil reconstruction}
\label{multicompare}
\begin{table}[t]
  \caption{\label{multicoil12xtab}The average MSE, PSNR, and SSIM of different methods at 12-fold acceleration on the multicoil cardiac cine test dataset (mean±std).}
  \centering
  \begin{tabular}{c|c|c|c}
  \hline\hline
  Methods &MSE(*e-5)	&PSNR	(dB)	&SSIM(*e-2)	\\
  \hline
  L+S &20.89±13.54	&37.52±2.35	&93.31±3.47\\
  MoDL	&6.84±4.02	&42.26±2.20	&96.84±1.24\\
  DL-ESPIRiT &4.27±2.03 &44.13±1.89 &97.79±0.83\\
  L+S-Net	&{\bf 2.67±1.34}	&{\bf 46.22±2.00}	&{\bf 98.49±0.60}\\
  

  \hline\hline
  \end{tabular}
\end{table}

Our proposed L+S-Net performs well on multicoil dynamic MR data. Because multicoil data contain more redundant information that can potentially improve the reconstruction quality, we started with an acceleration of 12. Fig.\ref{multicoil} shows the comparison between the results of L+S, MoDL, DL-ESPIRiT and the proposed L+S-Net at an acceleration factor of 12. L+S almost failed at such a high acceleration rate, and the reconstruction result shows many artifacts. The deep-learning-based MoDL and DL-ESPIRiT methods achieve acceptable results. We can see that the proposed L+S-Net performs better, especially from the error map. The quantitative results in Table~\ref{multicoil12xtab} correspond to the visual comparison.

It is worth noting that L+S-Net performs even better on 12-fold multicoil data than on 8-fold single-coil data due to the redundant information from multiple coils. However, the reconstruction of traditional L+S is not improved with the multicoil scheme. We assume that DL-based methods will benefit from the parallel imaging scheme more than traditional methods. Even with the help of the low-rank prior, the acceleration factor of 12 still exceeds the upper bound of the traditional L+S method.



\subsubsection{Prospective study on real-time OCMR data}
\label{prospectivecompare}
The free-breathing 9-fold undersampled cardiac dataset described in section \ref{prospectivedata}(2) was reconstructed using L+S, MoDL, DL-ESPIRiT and L+S-Net. The visual comparison is shown in Fig.~\ref{prospectivefig}. The real-time undersampled data were tested on the 12-fold model in a prospective study. Because the spatial resolution was lower than that of the data in the retrospective study, the quality of the images in this part was not as good as that in the retrospective study. The deep-learning-based methods are superior to the iterative L+S method, showing much lower streaking artifacts and spatial blurring. The results of MoDL and DL-ESPIRiT still have some artifacts around the red arrow, while the results of L+S-Net are cleaner.

\section{Discussion}
\subsection{Comparison of the reconstruction time}

The computation of singular value decomposition (SVD) is very time consuming, even when the reduced SVD is used. The iteration algorithms usually take dozens or hundreds of steps to converge, multiplying the time of SVD. Therefore, traditional low-rank methods in MRI reconstruction take much time to reconstruct. In our method, we unrolled the iteration into \emph{\textbf{ten}} steps, which means that only ten SVDs are computed in the reconstruction procedure. This significantly reduced the computation time. The average time used for reconstructing a dynamic series is listed in the last column in Table \ref{timetable}. The traditional iterative L+S method takes a much longer time than the other deep-learning-based methods, especially when reconstructing images from undersampled multicoil data. The proposed L+S-Net consumes slightly more time than the other deep learning methods compared in this work. Although the introduction of deep low-rank methods increases the amount of computation required for the SVD procedure in L+S-Net, the reconstruction time's effect is minimal compared to that of sparse-based deep learning methods and can even be ignored when compared with the reconstruction time of traditional CS methods.

\subsection{Very high acceleration reconstruction}
\begin{figure}[!h]
  \centering
  \includegraphics[width=\linewidth]{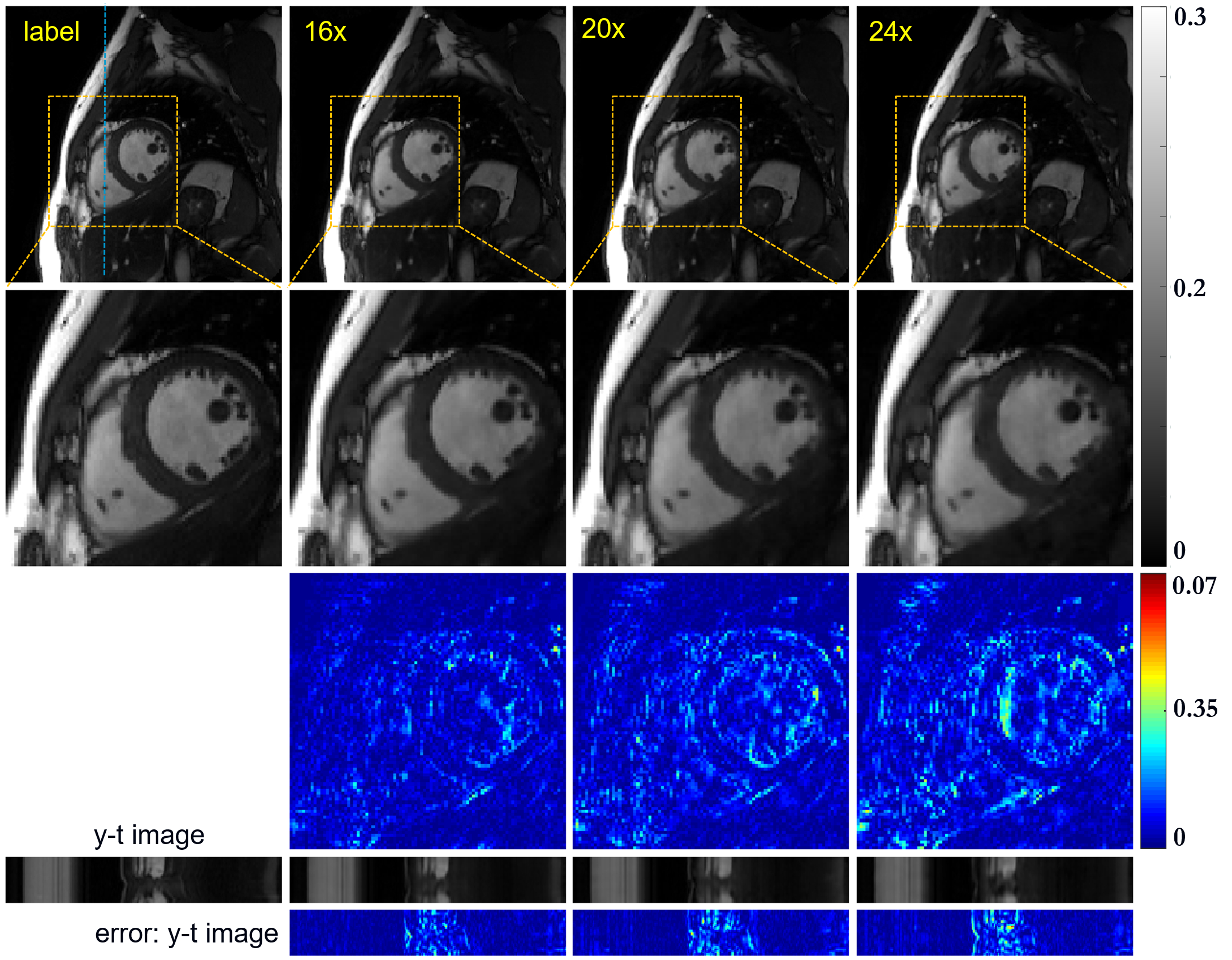}
\caption{\label{extremehighacc}The reconstruction results of the proposed L+S-Net at 16-fold, 20-fold and 24-fold acceleration on the multicoil cardiac cine dataset. From left to right, the first row shows the ground truth and the reconstruction result of these methods. The second row shows the enlarged views of the respective heart regions framed by a yellow box. The third row shows the error map (display range [0, 0.07]). The y-t image (extraction of the 92nd slice along the y and temporal dimensions) and the y-t image error are also given for each signal to show the reconstruction performance in the temporal dimension.
}
\end{figure}

\begin{table}[t]
  \caption{\label{extrehightab}The average MSE, PSNR and SSIM of L+S-Net at 16-fold, 20-fold and 24-fold acceleration on the multicoil cardiac cine test dataset (mean±std).}
  \centering
  \begin{tabular}{c|c|c|c}
  \hline\hline
  Acceleration &MSE(*e-5)	&PSNR	(dB)	&SSIM(*e-2)	\\
  \hline
  16x &3.94±2.01	&44.54±2.03	&97.89±0.80\\
  20x	&5.50±2.75	&43.09±2.02	&97.41±0.96\\
  24x	&7.16±3.41	&41.89±1.91	&96.73±1.09\\

  \hline\hline
  \end{tabular}
\end{table}
Because the multicoil reconstruction result of L+S-Net is still promising at 12-fold acceleration, we explored its performance at higher acceleration factors: 16-fold, 20-fold and 24-fold. The reconstruction results and error maps are shown in Fig. \ref{extremehighacc}. The details in both of the two results are acceptable. Our approach can still recover the image's dynamic content very well on such extremely undersampled data, although the artifacts near the boundary are slightly stronger at 24-fold. From the y-t view, the results of our proposed method are still acceptable for all three acceleration factors, but at 24-fold, the dynamic information loss is slightly larger than that at 16-fold and 20-fold. Table~\ref{extrehightab} shows the quantitative results. The reconstruction result at an extremely high acceleration rate indicates that the proposed method has great potential in high-resolution dynamic imaging. 

\subsection{Phase reconstruction}
\begin{figure}[!ht]
  \centering
  \includegraphics[width=\linewidth]{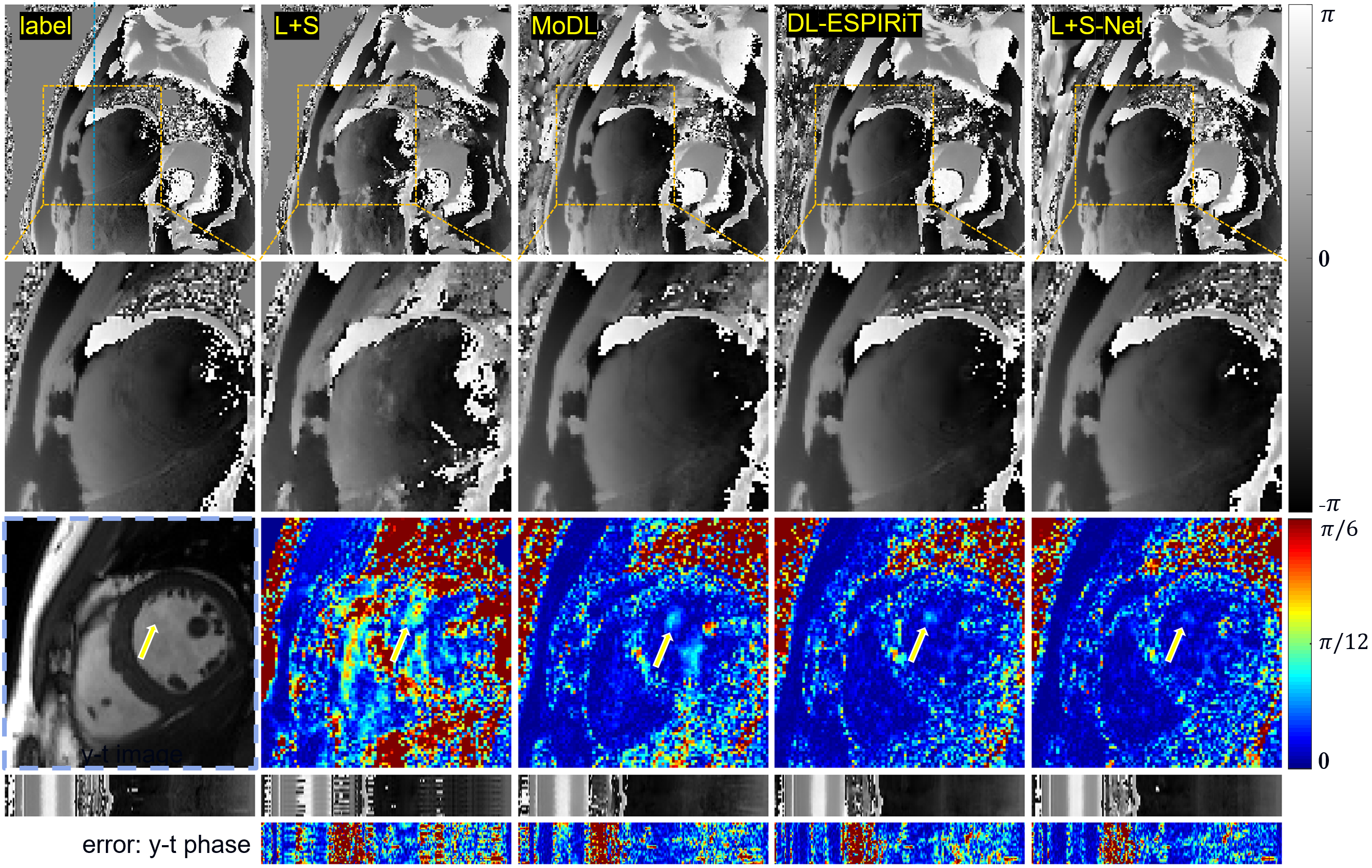}
\caption{\label{phase}
The phase reconstruction results of the different methods (L+S, MoDL, DL-ESPIRiT and L+S-Net) at 12-fold acceleration using multicoil data. From left to right, the first row shows the ground truth and the reconstruction results of these methods. The second row shows the enlarged views of their respective heart regions framed by a yellow box. The third row shows the reference magnitude image and error maps (display range [0, 0.07]). The y-t image (extraction of the 92nd slice along the y and temporal dimensions) and the y-t image error are also given for each signal to show the reconstruction performance in the temporal dimension.}
\end{figure}

Our model is designed for complex-valued MR data; therefore, it is also capable of reconstructing phase images accurately. Fig. \ref{phase} shows the phase that corresponds to each magnitude image in Fig. \ref{multicoil}. From left to right, the first row shows the ground truth and the reconstruction results of these methods. The second row shows the enlarged views of their respective heart regions framed by a yellow box. The third row shows the reference magnitude image and error maps (display range [0, 0.07]). The y-t image (extraction of the 92nd slice along the y and temporal dimensions) and the y-t image error are also given for each signal to show the reconstruction performance in the temporal dimension. Within the heart region, we can see that our method has the lowest error; the area near the yellow arrow is an example.

\subsection{Limitations of the proposed method}
Although the proposed L+S-Net achieved good reconstruction results, it still has some limitations. First, the scheme for learning the low-rank prior in our proposed method does not avoid the SVD calculation, which leads to a longer reconstruction time. LSVT is one of the many methods that can explore the low-rank priors. Further work needs to be done to explore whether a better low-rank prior scheme can be used in L+S-Net, such as methods involving the Hankel matrix, which can avoid the SVD calculation. Second, reconstruction is performed for a batch of frames each time, and reducing the amount of frames may degrade the effect of the low-rank prior. Therefore, it is not suitable for extremely low-latency imaging (for example, real-time navigation for interventional surgery). Third, the performance of L+S-Net is not tested on body parts other than the heart. More tests should be performed in different dynamic imaging scenarios.

\section{Conclusion and outlook}
In this paper, we proposed a model-based unrolled low-rank plus sparse network, dubbed L+S-Net, for dynamic MR reconstruction. In particular, we used an alternating linearized minimization method to solve the optimization problem with low-rank and sparse regularization. A learned singular value threshold was introduced to ensure the clear separation of the L component and S component. Then, the iterative steps were unrolled into a network whose regularization parameters are learnable. The convergence analysis guarantees the performance of the reconstruction theoretically. The proposed method was tested on two cardiac cine datasets. The experimental results showed that our method can improve the reconstruction results both qualitatively and quantitatively.

\section*{Acknowledgments}
This work was supported in part by the National Key Research and Development Program of China under Grant 2017YFC0108802 and Grant 2020YFA0712200;
in part by the National Natural Science Foundation of China under Grant 12026603, Grant U1805261,Grant 61771463, Grant 81971611, Grant 61871373, and Grant 81729003;
in part by the Key Field R\&D Program of Guangdong Province under Grant 2018B030335001; 
in part by the Science and Technology Plan Program of Guangzhou under Grant 202007030002;
in part by the Shenzhen Peacock Plan Team Program under Grant KQTD20180413181834876; 
in part by the Innovation and Technology Commission of the Government of Hong Kong SAR under Grant MRP/001/18X; 
in part by the Engineering Laboratory Program of Chinese Academy of Sciences under
Grant KFJ-PTXM-012;
in part by the Strategic Priority Research Program of Chinese Academy of Sciences under Grant XDC07040000 and Grant XDB25000000; 
and in part by the China Postdoctoral Science Foundation under Grant 2020M682990 and Grant 2021M69331.




\bibliographystyle{model2-names.bst}\biboptions{authoryear}
\bibliography{refs}



\end{document}


\verso{Wenqi Huang \textit{et~al.}}

{\leftline{\huge Supplementary Material}}
\section{Derivation from Eq. (4) to Eq. (5)}
\begin{equation}
  \label{LSmodel_my}
  \min_{L, S, X} \frac{1}{2}\| AX-y\|_2^2+\lambda_{L} \| L\|_{*} + 
  \mathcal{R}(S), \quad s.t. \quad X = L+S.
\end{equation}
The auxiliary is specifically described for the proposed L+S-Net. The introduction of the auxiliary variable $X$ enables an inexact search of $L$ and $S$ at the beginning of the iteration steps, making it faster to find the optimal $L$ and $S$. This is critical in unrolled deep learning methods. The number of iteration blocks in an unrolled network is much smaller than the number of iterations needed to converge in iterative methods. When unrolled into the network, the initial steps of the search play an important role in finding optimal points. The penalty function of Eq.\eqref{LSmodel_my} is:
\begin{equation}
  \label{mypenalty}
  J(X,L,S) = \frac{1}{2}\| AX-y\|_2^2+\lambda_{L} \| L\|_{*} + 
  \mathcal{R}(S) + \frac{\rho}{2}\|L+S-X\|_2^2,
\end{equation}
where $\frac{\rho}{2}$ is a penalty parameter. 

Eq.\eqref{mypenalty} can be solved via the following alternating minimization method
\begin{equation}
  \label{subproblemsshort}
  \left\{
  \begin{aligned}
    L_{k+1} &= \arg\min_{L} J(X_k, L, S_k) \\
    S_{k+1} &= \arg\min_{S} J(X_k, L_{k+1}, S) \\
    X_{k+1} &= \arg\min_{X} J(X, L_{k+1}, S_{k+1}) \\
  \end{aligned}     
  \right..
\end{equation}
Since the calculation of $(A^*A+\rho I)^{-1}$ is highly time consuming, we attempt to solve the $X$-subproblem inexactly. Specifically, for the $X$-subproblem in Eq.\eqref{subproblemsshort}, we can approximate the cost function $J$ by linearizing the data fidelity $F(X):= \frac{1}{2}\| Ax-y\|_2^2$ at $\hat{X}_{k+1} := L_{k+1} + S_{k+1}$, i.e., 
\begin{equation}
  X_{k+1} = \arg\min_{X} \tilde{J}_{k+1}(X, L_{k+1}, S_{k+1}), 
\end{equation}
where 
\begin{equation}
  \begin{aligned}
    \tilde{J}_{k+1}(X, L_{k+1}, S_{k+1}):= & <\nabla F(\hat{X}_{k+1}), X-\hat{X}_{k+1}> \\  
        &+ \frac{1}{2\eta}\|X-\hat{X}_{k+1}\|_2^2 + \lambda_{L}\|L_{k+1}\|_{*} \\
        &+\mathcal{R}(S_{k+1})+\frac{\rho}{2}\|L_{k+1}+S_{k+1}-X\|_2^2.
  \end{aligned}
\end{equation}
Thus, the subproblems in Eq.\eqref{subproblemsshort} can be reformulated as 
\begin{equation}
  \label{subproblemslong}
  \left\{
  \begin{aligned}
    L_{k+1} = \arg\min_{L} &\frac{\rho}{2}\|L+S_k-X_k\|_2^2 + \lambda_{L} \| L\|_{*}\\
    S_{k+1} = \arg\min_{S} &\frac{\rho}{2}\|L_{k+1}+S-X_k\|_2^2 + \mathcal{R}(S) \\
    X_{k+1} = \arg\min_{X} &\frac{\rho}{2}\|L_{k+1}+S_{k+1}-X\|_2^2 + 
              <\nabla F(\hat{X}_{k+1}), X> \\
              &+ \frac{1}{2\eta}\|X-\hat{X}_{k+1}\|_2^2\\
  \end{aligned}     
  \right..
\end{equation}

\section{Proof of Convergence}
If $f$ is $L$-Lipschitz continuous, then the following inequality holds for any $x,z$
$$|f(x)-f(z)|\leq L\|x-z\|.$$

If $f$ is $L$-smooth, then the gradient of $f$ is $L$-Lipschitz continuous. An extended-real-value function $f:dom(f)\rightarrow \mathbb{R}\cup\{+\infty\}$ is called proper if it is finite somewhere and never equals $-\infty$. We say $f$ is coercive if  $f(x)\rightarrow\infty$ holds as $\|x\|\rightarrow\infty$.

\begin{defn}
[KL function]\label{kl} A proper function $f(\cdot)$ is called a Kurdyka-{\L}ojasiewicz (KL) function if for any point $x$ where $f(\cdot)$ is subdifferentiable, i.e., $\partial f(x)\neq\varnothing$, it satisfies the KL inequality.
That is, for all $x'$  in the neighborhood of $x$, if $f(x)<f(x')<f(x)+\zeta$, the following inequality holds,
$$\varphi'(f(x')-f(x))\mathrm{dist}(0,\partial f(x'))\geq1,$$
where the scalar $\zeta\in[0,\infty)$, $\varphi:[0,\zeta)\rightarrow\mathbb{R}_{+}$ is a continuous concave function such that

\begin{itemize}
  \item $\varphi(0)=0$;

  \item $\varphi$ is continuous differentiable in $(0,\zeta)$;

  \item  for all $z\in(0,\zeta)$, $\varphi'(z)>0$.
\end{itemize}
\end{defn}

\begin{equation}
\widetilde{J}_{\rho,k}(L,S):=F(\widehat{X}_{k})+\langle \nabla F(\widehat{X}_{k}),L+S-\widehat{X}_{k}\rangle+\lambda_L\|L\|_*+\mathcal{R}_{\theta^{k+1}}(S)+\frac{\rho}{2}\|L+S-\widehat{X}_{k}\|^2
\end{equation}
\begin{equation}\label{adm}\left\{\begin{aligned}
L_{k+1} &= \arg\min_{L}\widetilde{J}_{k}(L,S_k)\\
S_{k+1} &= \arg\min_{S}\widetilde{J}_{k}(L_{k+1},S)\\
\end{aligned}\right.\end{equation}

\begin{assump}
\label{assupi}
~

(\textbf{A}1) $\mathcal{R}_{\theta^1}:\mathbb{C}^d\rightarrow \mathbb{R}\cup\{+\infty\}$ is a nonnegative, proper, coercive and $L_{\mathcal{R}_{\theta^1}}$-smooth KL function.

(\textbf{A}2) The sequence of paired proximal operators of $(\mathcal{R}_{\theta^k},\mathcal{R}_{\theta^{k+1}} )$, termed $(\mathcal{T}_{\theta^k},\mathcal{T}_{\theta^{k+1}})$, is asymptotically nonexpansive with a sequence $\{\epsilon^{k+1}\}$, i.e.,
$$\|\mathcal{T}_{\theta^k}(X)-\mathcal{T}_{\theta^{k+1}}(Y)\|^2\leq(1+\epsilon^{k+1}) \|X-Y\|^2,~~\forall X,Y,k.$$

\end{assump}
Assumptions (\textbf{A}1) and (\textbf{A}2) are standard assumptions for analyzing the convergence of learned iterative algorithms, which can also be found in \cite{8481558,9152164}.

In this subsection, we consider the convergence of Algorithm (\ref{adm}). Now, we define an auxiliary cost functional as follows:
\begin{equation*}
\widehat{J}_{\rho,k}(L,S):=F(\widehat{X}_{k})+\langle \nabla F(\widehat{X}_{k}),L+S-\widehat{X}_{k}\rangle+\lambda_L\|L\|_*+\mathcal{R}_{\theta^{1}}(S)+\frac{\rho}{2}\|L+S-\widehat{X}_{k}\|^2
\end{equation*}
and the objective is defined as follows:
\begin{equation}
J(L,S):=F(L+S)+\lambda_L\|L\|_*+\mathcal{R}_{\theta^{1}}(S).
\end{equation}

\begin{thm}
\label{thm1}
Suppose that Assumptions (\textbf{A}1)-(\textbf{A}2) hold and $\rho\geq\{L_{A^*A},\widetilde{\sigma}\}$, where $L_{A^*A}$ is the Lipschitz-continuous constant of $\nabla F$ and $\widetilde{\sigma}$ is the minimum constant such that $\mathcal{R}_{\theta^{1}}(\cdot) +\frac{\widetilde{\sigma}}{2}\|\cdot\|^2$ is $\sigma$+4-strongly convex. Then, the sequence $\{(L_{k},S_{k})\}$ generated by Algorithm (\ref{adm}) converges to a critical point of the objective $J(L,S)$.
\end{thm}
\begin{pf}
Before giving a concrete proof of theorem \ref{thm1}, we provide a useful lemma:
\begin{lem}
\label{thmi}
Supposed that Assumption \textbf{A}2 holds and $\rho\geq\{L_{A^*A},\widetilde{\sigma}\}$, where $L_{A^*A}$ is the Lipschitz-continuous constant of $\nabla F$ and $\widetilde{\sigma}$ is the minimum constant such that $\mathcal{R}_{\theta^{1}}(\cdot) +\frac{\widetilde{\sigma}}{2}\|\cdot\|^2$ is $\sigma$+4-strongly convex. For any $k\geq1$, the sequences $\{(x_{k},z_{k},w_k)\}$ generated by Algorithm \ref{adm} satisfy the following statements:

(\rmnum{1}) There exist two constants, i.e., $0<\mu_1\leq\frac{\rho}{2},~0<\mu_2 \leq \frac{\sigma}{2}$, such that
\begin{equation}
\begin{aligned}\widehat{J}_{\rho,k}(L_{k+1},S_k)+\mu_1\|L_{k+1}-L_{k}\|^2\leq \widehat{J}_{\rho,k}(L_{k},S_k),
\label{thmfo2}
\end{aligned}
\end{equation}

\begin{equation}
\begin{aligned}\widehat{J}_{\rho,k}(L_{k+1},S_{k+1})+\mu_2\|S_{k+1}-S_{k}\|^2\leq \widehat{J}_{\rho,k}(L_{k+1},S_k).
\label{thmfo2:2}
\end{aligned}
\end{equation}

(\rmnum{2})
\begin{equation}\label{thmfo3}
\lim_{k\rightarrow\infty}(\|L_{k+1}-L_{k}\|^2+\|S_{k+1}-S_{k}\|^2)=0.
\end{equation}

(\rmnum{3})
Let $\left(L^*_k,S^*_k\right)\in \partial J(L_{k},S_{k})$.
Then, for all bounded subsequences $\left\{\left(L_{k_i},S_{k_i}\right)\right\}$ of $\left\{\left(L_{k},S_{k}\right)\right\}$, $\left(L^*_{k_i},S^*_{k_i}\right)\rightarrow0$ as $i\rightarrow+\infty,$
i.e.,
 $$\mathrm{dist}\left(\partial J(L_{k_i},S_{k_i}),0\right)\rightarrow 0 ~\text{when}~i\rightarrow+\infty.$$
\end{lem}
\begin{pf}
(\rmnum{1}) Inequality (\ref{thmfo2:2}) can be obtained by direct calculation. Now, we present the proof of inequality (\ref{thmfo2}). Since $\widehat{J}_{\rho,k}(X,L,S)$ is $\sigma$+4-strongly convex with respect to $S$, we have
$$\widehat{J}_{\rho,k}(L_{k+1},S_k)-\widehat{J}_{\rho,k}(L_{k+1},S_{k+1})-\langle\nabla_{S_{k+1}}\widehat{J}_{\rho,k}(L_{k+1},S_{k+1}),S_{k}-S_{k+1} \rangle\geq\frac{\sigma+4}{2}\|S_{k+1}-S_{k} \|^2.$$
By the definition of $\widehat{J}_{\rho,k}(L,S)$, $\nabla_{S_{k+1}}\widehat{J}_{\rho,k}(L_{k+1},S_{k+1})=\rho(\mathcal{T}_{\theta^{k}}(S'_{k+1})-S'_{k+1})+\nabla \mathcal{R}_{\theta_1}(\mathcal{T}_{\theta^{k}}(S'_{k+1}))$ holds, where $ S'_{k+1}:=\widehat{X}_{k}-\frac{1}{\rho}F(\widehat{X}_{k})-L_{k+1}$. Then, we have
\begin{equation}\label{ineq:a}\begin{aligned}
&\widehat{J}_{\rho,k}(L_{k+1},S_k)-\widehat{J}_{\rho,k}(L_{k+1},S_{k+1})\\
=&\rho\langle \mathcal{T}_{\theta^{k}}(S'_{k+1})- \mathcal{T}_{\theta^{1}}(S'_{k+1}),S_{k+1}-S_k\rangle+\frac{\sigma+4}{2}\|S_{k+1}-S_{k} \|^2\\
&+\langle \nabla \mathcal{R}_{\theta_1}(\mathcal{T}_{\theta^{k}}(S'_{k+1}))- \nabla \mathcal{R}_{\theta_1}(\mathcal{T}_{\theta^{1}}(S'_{k+1})),S_{k+1}-S_k\rangle\\
\geq&-2(\rho+L_{\mathcal{R}_{\theta_1}})^2\|\mathcal{T}_{\theta^1}(S'_{k+1})- \mathcal{T}_{\theta^{k}}(S'_{k+1})\|^2+\frac{\sigma}{2}\|S_{k+1}-S_k\|^2\\
\geq&-2(\rho+L_{\mathcal{R}_{\theta_1}})^2\|\sum_{i=1}^{k-1}\mathcal{T}_{\theta^i}(S'_{k+1})- \mathcal{T}_{\theta^{i+1}}(S'_{k+1})\|^2+\frac{\sigma}{2}\|S_{k+1}-S_k\|^2\\
\geq&-2(\rho+L_{\mathcal{R}_{\theta_1}})^2k\sum_{i=1}^n\|\mathcal{T}_{\theta^i}(S'_{k+1})- \mathcal{T}_{\theta^{i+1}}(S'_{k+1})\|^2+\frac{\sigma}{2}\|S_{k+1}-S_k\|^2\\
\geq&-2(\rho+L_{\mathcal{R}_{\theta_1}})^2k\sum_{i=1}^n(1+\epsilon^k)\cdot0+\frac{\sigma}{2}\|S_{k+1}-S_k\|^2\\
=&\frac{\sigma}{2}\|S_{k+1}-S_k\|^2
\end{aligned}\end{equation}
where the first equality is due to the relation $\rho(\mathcal{T}_{\theta^{1}}(S'_{k+1})-X_k+L_{k+1})+\nabla \mathcal{R}_{\theta_1}(\mathcal{T}_{\theta^{1}}(S'_{k+1}))=0$, which is derived by the first-order optimization condition for $\widehat{J}_{\rho,k}(L_{k+1},S_{k+1})$ at $S_{k+1}$; the second inequality is due to $\langle a,b\rangle\leq2\|a\|^2+2\|b\|^2$; and the last inequality makes use of Assumption (\textbf{A}2).

For (\rmnum{2}), according to inequalities (\ref{thmfo2}) and (\ref{thmfo2:2}), we have
\begin{equation*}\widehat{J}_{\rho,k}(L_{k},S_k)-\widehat{J}_{\rho,k}(L_{k+1},S_{k+1})\geq\frac{\rho}{2}\|L_{k+1}-L_{k}\|^2+\frac{\sigma}{2}\|S_{k+1}-S_{k}\|^2.
\end{equation*}
Moreover, we have
\begin{equation}\label{ineq:12}\begin{aligned}
&\widehat{J}_{\rho,k}(L_{k+1},S_{k+1})-\widehat{J}_{\rho,k+1}(L_{k+1},S_{k+1})\\
=&F(\widehat{X}_{k})-F(\widehat{X}_{k+1})+\langle\nabla F(\widehat{X}_{k}),\widehat{X}_{k+1}-\widehat{X}_k \rangle+\frac{\rho}{2}\|\widehat{X}_{k+1}-\widehat{X}_k\|^2\\
\geq&\left(\frac{\rho}{2}-\frac{L_{A^*A}}{2}\right)\|\widehat{X}_{k+1}-\widehat{X}_k\|^2\geq0\\
\end{aligned}\end{equation}
where the  inequality makes use of the $L_{A^*A}$-Lipschitz continuity of $\nabla{F}$.
Then, we have
\begin{equation*}
\mu(\|L_{k+1}-L_{k}\|^2+\|S_{k+1}-S_{k}\|^2)\leq\widehat{J}_{\rho,k}(L_{k},S_k)-\widehat{J}_{\rho,k+1}(L_{k+1},S_{k+1}),
\end{equation*}
where $\mu\leq\frac{1}{2}\min\{\rho,\sigma\}$. Summing the above inequality from 0 to $\infty$ yields
\begin{equation*}
\mu\sum_{k=0}^{+\infty}(\|L_{k+1}-L_{k}\|^2+\|S_{k+1}-S_{k}\|^2)\leq\widehat{J}_{\rho,0}(L_{0},S_0)<+\infty.
\end{equation*}
Then, we obtain (\ref{thmfo3}).

(\rmnum{3}) By the definition  of $(L_{k+1}^*,S_{k+1}^*)\in\partial J(L_{k+1},S_{k+1})$, we have
 \begin{equation*}
\left\{\begin{aligned}
L_{k+1}^*\in&\nabla F(\widehat{X}_{k+1})+\lambda_{L} \partial\|L_{k+1}\|_*,\\
S_{k+1}^*=&\nabla F(\widehat{X}_{k+1})+\nabla \mathcal{R}_{\theta^1}(S_{k+1})\\
\end{aligned}\right.
\end{equation*}
Furthermore, by the iterative rule of Algorithm (\ref{adm}), we have
$$0\in\rho(L_{k+1}-L_k)+\nabla F(\widehat{X}_{k})+\lambda_{L} \partial\|L_{k+1}\|_*.$$
Then, we obtain
\begin{equation}\label{equ:l}\begin{aligned}
\|L_{k+1}^*\|=&\|\rho(L_{k+1}-L_k)+\nabla F(\widehat{X}_{k})-\nabla F(\widehat{X}_{k+1})\|\\
\leq&\rho\|L_{k+1}-L_k\|+L_{A^*A}\|\widehat{X}_{k}-\widehat{X}_{k+1}\|
\end{aligned}\end{equation}
Following (\ref{thmfo3}), it is easy to verify $\lim_{k\rightarrow+\infty} L_{k}^*=0$.
According to the first-order condition of Algorithm (\ref{adm}) and inequality (\ref{ineq:a}), we know that
$$\rho(\widehat{X}_{k+1}-\widehat{X}_{k})+\nabla F(\widehat{X}_{k})+\nabla \mathcal{R}_{\theta^{1}}(S_{k+1})=0.$$
Thus, following the same derivation approach, we have
\begin{equation}\label{equ:s}\begin{aligned}
\|S_{k+1}^*\|=&\|\rho(\widehat{X}_{k+1}-\widehat{X}_k)+\nabla F(\widehat{X}_{k})-\nabla F(\widehat{X}_{k+1})\|\\
\leq&(\rho+L_{A^*A})\|\widehat{X}_{k}-\widehat{X}_{k+1}\|
\end{aligned}\end{equation}
which verifies $\lim_{k\rightarrow+\infty}S_{k+1}^*=0$.
\end{pf}

Now, we prove the result of Theorem \ref{thm1}.
Theorem 2.9 of \cite{AttouchH2005} shows that the convergence of Algorithm (\ref{adm}) can be derived by verifying only the following three conditions.

(\textbf{C}1) (Sufficient decrease condition). For some $a>0$, it holds that, for each $k>0$,
\begin{equation*}
J(L_k,S_k)-J(L_{k+1},S_{k+1})\geq a(\|L_{k+1}-L_{k}\|^2+\|S_{k+1}-S_{k}\|^2).\\
\end{equation*}

(\textbf{C}2) (Relative error condition). For some $b>0$, there exist some
$\left(L^*_{k+1},S^*_{k+1}\right)\in\partial J(L_{k+1},S_{k+1})$ such that
\begin{equation*}
\left\|\left(L^*_{k+1},S^*_{k+1}\right)\right\|\leq b(\left\|L_{k+1}-L_{k}\right\|+\left\|S_{k+1}-S_{k}\right\|).
\end{equation*}

(\textbf{C}3) (Continuity condition). There exists a subsequence $
\{(L_{k_i},S_{k_i})\}$ of the sequence generated by Algorithm (\ref{adm}) and a cluster point $\left(\mathring{L},\mathring{S}\right)$ such that $\left\{(L_{k_i},S_{k_i})\right\}\rightarrow\left(\mathring{L},\mathring{S}\right)$ and $J(L_{k_i},S_{k_i})\rightarrow J\left(\mathring{L},\mathring{S}\right)$, as $i\rightarrow+\infty$.

(\textbf{C}1) can be derived directly from (\ref{thmfo2}), (\ref{thmfo2:2}) and the relation $J(L_{k},S_{k})=\widehat{J}_{\rho,k}(L_{k},S_{k})$, and (\textbf{C}2) can be deduced directly from (\ref{equ:l}) and (\ref{equ:s}). Now, we give the proof for condition (\textbf{C}3).

On the basis of Assumptions (\textbf{A}1) and (\textbf{A}2), we know that $ J(L,S)$ is proper l.s.c. and coercive with respect to $L$ and $S$. Hence, $\{L_k,S_k\}$ is contained in the level set $\{(L_k,S_k)|~ J(L_k,S_k)\leq J(L_0,S_0)\}$. Using the Bolzano-Weierstrass theorem, we deduce the existence of a subsequence, denoted as $\{L_{k_i},S_{k_i}\}$, that converges to some cluster point $\left\{\mathring{L},\mathring{S}\right\}$.
Note from the definition of $S_{k_i+1}$ as a minimizer that
\begin{equation*}
\begin{aligned} \widehat{J}_{\rho,k_i}(L_{k_i+1},S_{k_i+1})\leq \widehat{J}_{\rho,k_i}(L_{k_i+1},\mathring{S}).\end{aligned}
\end{equation*}
Following (\ref{ineq:12}), we also have
\begin{equation*}
\begin{aligned} \widehat{J}_{\rho,{k_i+1}}&(L_{k_i+1},S_{k_i+1})+\left(\frac{\rho}{2}-\frac{L_{A^*A}}{2}\right)\|\widehat{X}_{k_i+1}-\widehat{X}_{k_i}\|^2\\
&\leq \widehat{J}_{\rho,{k_i}}(L_{k_i+1},S_{k_i+1}),\end{aligned}
\end{equation*}
Based on the relation $\widehat{J}_{\rho,k_i+1}(L_{k_i+1},S_{k_i+1})=J(L_{k_i+1},S_{k_i+1})$, we obtain
\begin{equation}\label{limjk}
\begin{aligned} \limsup_{i\rightarrow+\infty}& \left\{\widehat{J}_{\rho,{k_i+1}}(L_{k_i+1},S_{k_i+1})+\left(\frac{\rho}{2}-\frac{L_{A^*A}}{2}\right)\|\widehat{X}_{k_i+1}-\widehat{X}_{k_i}\|^2   \right\}\\
&\leq\lim_{i\rightarrow+\infty} \widehat{J}_{\rho,k_i}(L_{k_i+1},\mathring{S}).\end{aligned}
\end{equation}
Since $ \{L_{k_i},S_{k_i}\}$ converges to $\left(\mathring{L},\mathring{S}\right)$, it is easy verify that $\{L_{k_i+1},S_{k_i+1}\}$ also converges to $\left(\mathring{L},\mathring{S}\right)$ according to inequality (\ref{thmfo3}) in Lemma \ref{thmi}. By the continuity of $\mathcal{R}_{\theta^1}$ and $\|\cdot\|_*$, we also have
\begin{equation*}
\lim_{i\rightarrow+\infty} \widehat{J}_{\rho,k_i}(L_{k_i+1},\mathring{S})=J(\mathring{L},\mathring{S}).
\end{equation*}
Then, we have
$$\limsup_{i\rightarrow+\infty}J(L_{k_i+1},S_{k_i+1})\leq J(\mathring{L},\mathring{S}).$$
Furthermore, the lower semicontinuity of $J(\cdot)$ implies that
\begin{equation*}
\liminf_{i\rightarrow+\infty} J(L_{k_i+1},S_{k_i+1})\geq J(\mathring{L},\mathring{S}).
\end{equation*}
Then, we prove that there exists a subsequence $\{L_{k_i+1},S_{k_i+1}\}$ that converges to $\left(\mathring{L},\mathring{S}\right)$ with  $J(L_{k_i+1},S_{k_i+1})\rightarrow J(\mathring{L},\mathring{S})$ as $i\rightarrow+\infty$. Then, (\textbf{C}3) is obtained.

Since conditions (\textbf{C}1), (\textbf{C}2) and (\textbf{C}3) hold, we have
$\{(L_{k},S_{k})\}$ converges to $\left(\mathring{L},\mathring{S}\right)$ following the result of Theorem 2.9 of \cite{AttouchH2005} directly.
Finally, according to the result (\rmnum{3}) in Lemma \ref{thmi}, it is easy to see that $\left(\mathring{L},\mathring{S}\right)$ is a critical point of $J(L,S)$. This concludes the proof.
\end{pf}


\bibliographystyle{model2-names.bst}\biboptions{authoryear}
\bibliography{supp_refs}